\documentclass[journal=jctcce,manuscript=article]{achemso}
\usepackage{amsmath}
\usepackage{chemformula} 
\usepackage[T1]{fontenc} 
\usepackage{graphicx}
\usepackage{caption}
\usepackage{subcaption}
\usepackage{dcolumn}
\usepackage{bm}
\usepackage{svg}
\usepackage{amsmath}
\usepackage{graphicx}
\usepackage{subcaption}
\usepackage{xcolor, soul} 
\usepackage{braket} 
\usepackage{amsmath}
\usepackage{amsfonts}
\usepackage{amssymb}
\usepackage{bbold}
\usepackage{siunitx}

\usepackage{xcolor}
\usepackage{soul}

\usepackage[]{changes}
\usepackage{xcolor}
\usepackage{tabularray}
\usepackage[version=4]{mhchem}
\usepackage{textgreek}
\usepackage{braket}
\usepackage{hyperref}
\usepackage[utf8]{inputenc}
\usepackage[T1]{fontenc}
\usepackage{mathptmx}
\usepackage{amsmath}
\usepackage{etoolbox}

\author{Mohammad Shams}
\author{Kimia Kargar}
\affiliation{Chemistry Department, Sharif University of Technology, Tehran 11155-9516, Iran}

\author{David Mendive-Tapia}
\affiliation[Heidelberg University]
{Theoretical Chemistry, Institute of Physical Chemistry, Heidelberg University, Im Neuenheimer Feld 229, 69120 Heidelberg, Germany}
\email {david.mendive-tapia@pci.uni-heidelberg.de}

\author{Fatemeh Khalili}
\affiliation{Chemistry Department, Sharif University of Technology, Tehran 11155-9516, Iran}

\author{Oriol Vendrell}
\affiliation[Heidelberg University]
{Theoretical Chemistry, Institute of Physical Chemistry, Heidelberg University, Im Neuenheimer Feld 229, 69120 Heidelberg, Germany}
\email {oriol.vendrell@uni-heidelberg.de}

\author{Zahra Jamshidi}
\email {njamshidi@sharif.edu}
\affiliation{Chemistry Department, Sharif University of Technology, Tehran 11155-9516, Iran}

\title{\Large Plasmonic Cavity Quantum Dynamics under Linear Vibronic Coupling}

\begin{tocentry}
\centering
\includegraphics[width=0.88\linewidth]{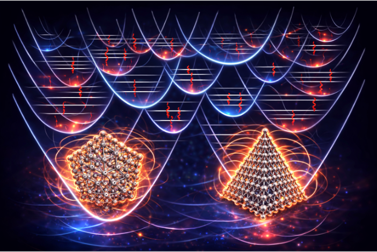}
\end{tocentry}

\begin{document}

\begin{abstract}
Modeling the quantum dynamics of plasmonic excitations—collective oscillations
    of free electrons interacting with light—remains a significant theoretical
    challenge, particularly due to the need to accurately describe their quantum
    nature and the role of non-radiative decay channels.  At the same time, a
    reliable theoretical framework is essential for advancing applications
    ranging from materials design to the development of new quantum optical
    platforms for quantum technologies.  In this work, we address these
    challenges by introducing a Hermitian formalism based on the linear vibronic
    coupling (LVC) model for the description of plasmonic excitations in
    metallic nanostructures.  This is parameterized through first-principles
    calculations --- including but not limited to, the full DFT ground state
    with tight-binding excited states --- and machine learning techniques using
    a newly implemented automated platform named Python Plasmonic Cavity (PyPC).
    The effectiveness of this workflow is demonstrated by successfully
    reproducing the experimental absorption spectra and vibronic broadening of
    plasmonic silver nanoparticles containing more than a hundred atoms.
    Additionally, the population dynamics of plasmonic states are investigated,
    showing that the LVC model accurately predicts ultrafast lifetimes for
    bright states and effectively captures the dynamics of dark states.

\end{abstract}

\section{Introduction}

Plasmonic excitations are the collective response of free electrons to an external electromagnetic field (EMF) such as light. They exhibit strong and tunable absorption that can focus the EMF in sub-diffraction limit volumes, leading to a wide range of applications\cite{maier2007localized}. In this context, a plasmonic nanocavity (PNC) with a well-confined EMF, enables robust light-matter interactions with a few or even single emitters at room temperature.\cite{leng2018strong,li2022room,bitton2020vacuum} This strong interaction between the EMF and nearby molecules leads to enhanced absorption or emission, as well as surface-enhanced infrared and Raman spectroscopy. \cite{neubrech2017surface,schlucker2014surface,bauch2014plasmon} The broad applicability of these phenomena makes plasmonic nanoparticles attractive for quantum optical applications,\cite{xu2018quantum,you2020multiparticle} and has prompted numerous studies in the field of polaritonics, with the goal of controlling these interactions by exploring how plasmonic nanomaterials interact with individual emitters\cite{xiang2024molecular,collins2020plasmon,yoshie2004vacuum}.

Nonetheless, the theoretical modeling of PNCs and their interaction with emitters presents several challenges. First of all, there is the problem of accounting for the dissipative nature of PNCs due to non-radiative decay channels when estimating cavity lifetime and performance. In short, although the extremely compact nature and intense electromagnetic fields of PNCs enable effective coupling even under ambient conditions,\cite{hugall2018plasmonic} they typically suffer from short lifetimes and moderate quality factors due to intrinsic non-radiative losses. These characteristics depend on various experimental factors, such as the quantum efficiency of the cavity, the photophysical properties of the emitters, and the distance and orientation of the emitters relative to the cavity.\cite{matsuzaki2021quantum}
Second, incorporating quantum features into the description of plasmonic properties is crucial in order to model the ultra-intense plasmonic fields and nonlinear optical properties of coupled plasmon-emitter systems.\citep{xu2018quantum,marinica2012quantum,zhong2020nonlinear} However, the quantum mechanical treatment of plasmonic nanoparticles has relevant limitations, as it needs to take into account the strong electron correlations and relativistic effects across numerous electronic states; which are confined within a narrow range of energy. Furthermore, these states are intertwined with the vibrational bath and exhibit strong coupling to the nuclear degrees of freedom (DOF)\cite{shvartsburg2001dissociation}. 

How can both of these issues be addressed in a practical manner?
In this regard, the linear vibronic coupling (LVC) Hamiltonian provides a conceptually simple and easily parameterizable framework that can effectively bridge the significant disparity in length scales between simulations of PNCs and interacting molecules by describing them on the same footing. In particular, the LVC Hamiltonian can capture the vibronic structure of PNCs of medium size, up to about 100 atoms, where a bulk description is not sufficiently accurate, and where an atomistic description of their electronic states and vibrational modes is still relevant to reproduce their interaction with surrounding molecules.
By doing so, it circumvents the need for multiscale modeling approaches, such as QM/MM.
Furthermore, the fact that the basis of vibronic coupling models lies in the use of a Taylor series expansion of the potential and couplings with respect to vibrational coordinates, i.e., nuclear displacements, allows for a set of increasingly improved descriptions of the interacting molecules; starting from the LVC as the reference framework.
%
For instance, this type of modeling has already been efficiently applied in several studies,\cite{zauleck2016two,harabuchi2016exploring,zobel2021quest} including tens of electronic states (up to 20) and up to approximately 100 vibrational degrees of freedom.\cite{yaghoubi2020ultrafast,aranda2021vibronic,harabuchi2016exploring} Recently, LVC modelling was combined with trajectory surface hopping (TSH) within the framework of the SHARC (surface hopping including arbitrary couplings) package.\cite{polonius2023lvc,mai2018nonadiabatic,zobel2021surface}

Therefore in this work we present the Python Plasmonic Cavity (PyPC) tools, a Python-based platform that automates the parametrization of Hamiltonian terms from electronic structure packages,
for a general linear vibronic coupling model in the context of quantum dynamics simulations in PNCs. PyPC generates input and operator files compatible with the Heidelberg package \cite{worth2000heidelberg,beck2000multiconfiguration}, and supports quantum chemistry packages such as the Amsterdam Modelling Suite (AMS) \cite{ADF2001}, ORCA \cite{neese2012orca,neese2018software}, and Q-Chem \cite{epifanovsky2021software}. The PyPC code is available in our GitHub repository, with the link provided in the Supporting Information.
%
All the necessary quantities are derived from ab initio computations of the ground state Hessian and the excited-state electronic energies, gradients, and non-adiabatic couplings (NACs) at this geometry.
This model is subsequently propagated using the multilayer multiconfiguration time-dependent Hartree (ML-MCTDH) approach, \citep{vendrell2011multilayer} enabling quantum dynamics simulations for systems ranging from a few atoms to over a hundred.

For small clusters, vibronic coupling terms can be directly computed  from first-principles (FP) calculations. This approach is practical for metal clusters of up to $\sim$20 atoms, however, for systems containing $\sim$100 atoms or more, direct computation becomes impractical due to prohibitive computational costs. Nonetheless, thanks to recent advances in machine learning --- which provide a practical alternative\citep{Westermayr2020} --- the PyPC tools also integrate several data-driven models with generative machine-learning methods, to efficiently replace the expensive electronic-structure FP calculations for large systems; within this newly unified framework minimal effort is required to parametrize LVC Hamiltonians with many nuclear DOFs that otherwise would not be possible, thus significantly enhancing the efficiency of quantum dynamics modeling of PNCs with large metal clusters.

\section{Methods}

\subsection{The Linear Vibronic Coupling Hamiltonian}

In order to avoid singularities in the non-adiabatic coupling terms, we represent the molecular Hamiltonian in a diabatic electronic basis. The electronic diabatic Hamiltonian ($\widehat{\mathbf{W}}_{n,m}$) can then be expressed as a Taylor series expansion of the diabatic potential in terms of dimensionless normal coordinates, $\mathbf{Q}$, around the ground-state equilibrium geometry, $\mathbf{Q}_0$,
and a basis of electronic states, $\{\ket{n}\}$:\cite{kouppel1984multimode}
\begin{equation}
\label{eqn:eq1}
\small
\widehat{\mathbf{W}}_{n,m}= \\
\sum_{i=1}^{3N-6}\frac{ \omega_{i}}{2}(\frac{\partial^2 }{\partial Q_i^2}+\hat{Q}_{i}^{2}) \mathbb{1} + \\ 
E_{n} \ket{n}\bra{n} + \\ 
\sum_{i}^{3N-6}\kappa_{i}^{(n)} Q_{i} \ket{n}\bra{n} + \\ 
\sum_{i}^{3N-6}\lambda_{i}^{(n,m)} Q_{i} \ket{n}\bra{m}, \\ 
\end{equation}
where the expansion around $\mathbf{Q}_0$ yields the reference zeroth-order Hamiltonian in the harmonic approximation as the first term. Here $\omega_i$ is the $i$-th frequency of the vibrational mode $Q_i$ and  $\mathbb{1}$ represents the unit matrix in the space of electronic states. The second summation expresses the energy shifts, in which $E_{n}$ is the vertical excitation energy from the ground, $n=0$, to the $n$-th excited electronic state. Last, $\kappa_{i}^{(n)}$ are the intrastate coupling and $\lambda_{i}^{(n,m)}$ (with $n \neq m$) are the non-adiabatic interstate coupling terms.

The corresponding dimensionless normal coordinates in terms of displacements from the reference point, $\mathbf{Q}_0$, in Cartesian coordinates is given by: 
\begin{equation}
\label{eqn:eq2}
Q_i=\sqrt{\frac{\omega _i}{\hbar }}\sum_{\alpha}T_{\alpha i}{M_{\alpha}^{1/2}}r_{\alpha}
\end{equation}

where $r_{\alpha}$ is the Cartesian coordinate, $\mathbf{T}_{i}$ is the ${i}$-th normal mode vector in mass-weighted Cartesian coordinates and $\mathbf{M}$ is diagonal matrix of atomic masses. The first-order intrastate couplings describe the shifts of the minima in the excited state potential energy surfaces with respect to the ground state at the Franck-Condon point, while the interstate couplings determine the interaction between the $n$-th and $m$-th electronic states under first-order displacements of the $Q_i$ mode.

\begin{equation}
\label{eqn:eq3}
\kappa_i^{(n)}=\sqrt{\frac{\hbar }{\omega_{i}}}\sum_{\alpha}\frac{\partial E_{n}}{\partial r_{\alpha}}\frac{T_{\alpha i}}{\sqrt{M_{\alpha}}}=\left. {\frac{\partial E_n}{\partial {Q}_i}} \right|_{\mathbf{Q}=\mathbf{Q}_0}
\end{equation}

\begin{equation}
\label{eqn:eq4}
\lambda_i^{(n,m)}=\sqrt{\frac{\hbar }{\omega _{i}}}\sum_{\alpha}\frac{\partial \left\langle \Psi_{n}|\widehat{H}|\Psi_{m} \right\rangle}{\partial {r}_\alpha}\frac{T_{\alpha i}}{\sqrt{M_{\alpha}}}= \left.{\frac{\partial \left\langle \Psi_{n}|\widehat{H}|\Psi_{m} \right\rangle}{\partial {Q}_i}} \right|_{\mathbf{Q}=\mathbf{Q}_0} 
\end{equation}

Tetrahedral silver clusters contain symmetry-induced Jahn–Teller (JT) crossings of the type $T \otimes  (e + t_2)$ , i.e., a triply degenerate electronic state ($T_{2}$) coupled to both doubly degenerate (e) and triply degenerate ($t_2$) vibrational modes. Applying the elementary symmetry selection rule, $\Gamma_{n} \otimes \Gamma_{m} \otimes \Gamma_{i} \supset A$, these electronic states go under first-order JT splitting when distorted along either the doubly degenerate (e) or the triply degenerate ($t_2$) vibrational modes. The totally symmetric ($a_1$) vibrational modes do not split the degeneracy as they shift all three states equally and preserve the tetrahedral symmetry of the cluster.

Therefore, it should be noted that in Eq. \ref{eqn:eq1} the totally symmetric $a_1$ vibrational modes (non-JT) appear only in the diagonals $\kappa_i^{(n)}$, 
while $t_2$ modes (JT-active) appear only in the off-diagonals and $e$ modes (JT-active) 
are present in both diagonals and off-diagonals. The latter, doubly-degenerate $e$ modes, fulfil the additional constraint that tunning ($\theta$) and coupling ($\epsilon$) modes satisfy the relation $\kappa_\theta^{(n)} = -\kappa_\theta^{(m)} = \lambda_{\epsilon}^{(n,m)}$. 
Similarly, additional intra-state coefficients and pseudo-JT inter-state couplings that do not meet the latter equality can be present due to a $T_{2}$ state interacting nearby a $E$ or a $T_{1}$ electronic state.  Moreover, due to selection rules, $T_{2}$ states are also dipole-allowed; we will refer to them as bright states (BS), while the remaining possible symmetries . i.e. $T_{1}$, $E$, $A_2$ and $A_1$, as dark states (DS).

\subsection{Generative regression by conditional density estimation (GRCDE)}

\begin{figure}[ht!]
  \centering
  \includegraphics[width=0.50\textwidth]{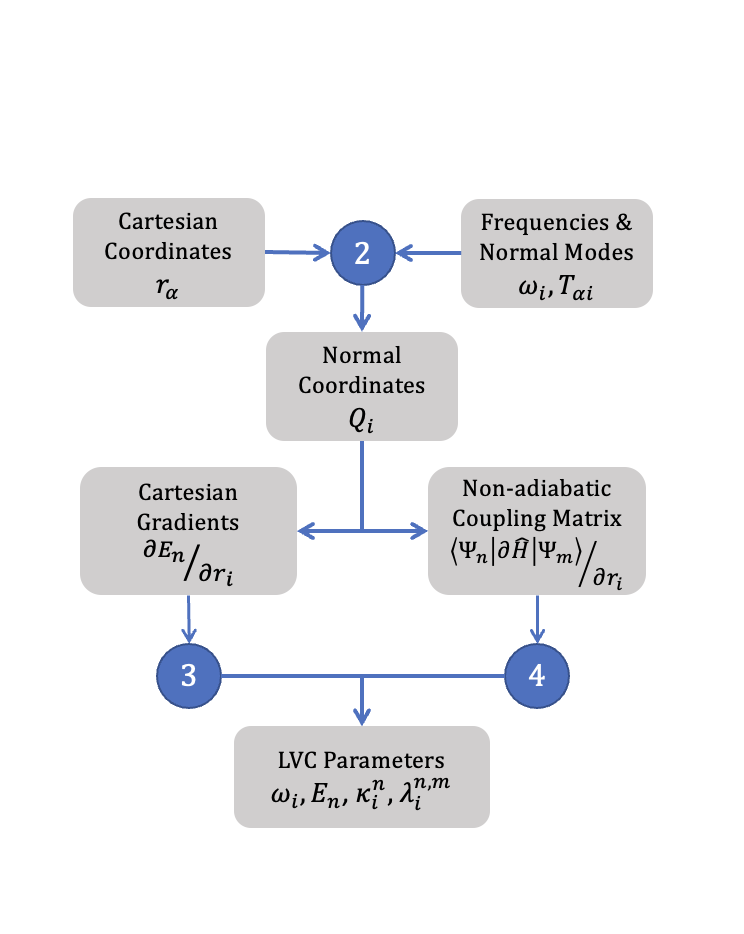}
  \caption{Generate LVC parameters based on FP calculations. Blue circles refer to the equation numbers.}
  \label{fig:fig1}
\end{figure}

Depending on the size of the system, two strategies can be employed to construct a LVC Hamiltonian.  
For "small clusters", the vibronic coupling terms are obtained directly from first‑principles (FP) calculations (Fig.\ref{fig:fig1} and Eqs.\ref{eqn:eq2}–\ref{eqn:eq4}).  
This procedure requires the Cartesian geometry, the harmonic vibrational frequencies, and the normal‑mode vectors in order to build the dimensionless normal coordinates (Eq. \ref{eqn:eq2}).  
Energy gradients and non‑adiabatic couplings are then projected onto this coordinate system, yielding the intrastate couplings \(\kappa_i^{(n)}\) and the interstate couplings \(\lambda_i^{(n,m)}\) (Eqs. \ref{eqn:eq3} and \ref{eqn:eq4}).  

In systems containing $\sim100$ atoms or more, a "second approach" is used.  
Here the LVC Hamiltonian is parametrised rapidly by training an instance‑based predictor on a reference data set \(\mathbf D\) that has been generated from FP calculations on smaller, tractable systems.
The predictor --- implemented via "generative regression by conditional density estimation (GRCDE)" --- exploits the information contained in the small systems to infer any desired quantity, in our case the diabatic coupling terms \(\kappa_i^{(n)}\) and \(\lambda_i^{(n,m)}\), for a much larger target system.

To understand the matrix \(\mathbf D\) that represents the dataset, consider one row at a time. Each row aggregates information from both the electronic and the vibrational degrees of freedom. As a concrete example, let the reference system be the \(\mathrm{Ag}_{10}\) cluster. For this system we have (i) the energies of nine low‑lying excited electronic states, (ii) the harmonic vibrational frequencies and normal-mode symmetries at the equilibrium geometry, and (iii) at least one of the two quantities required to recover the LVC parameters: either an excited-state energy gradient (for \(\kappa_i^{(n)}\)) or a non-adiabatic coupling vector (for \(\lambda_i^{(n,m)}\)).  
All of these data are combined into a single table. Each electronic state can be coupled to the \(3N-6\) normal modes.  Consequently, \(\mathrm{Ag}_{10}\) cluster with nine electronic states and 24 vibrational degrees of freedom yields \(9\times24 = 216\) rows, each of which we call a "sample".

The electronic and vibrational information form the "features".  
For the intrastate coupling \(\kappa_i^{(n)}\) the feature vector consists of the energy of the \(n\)-th electronic state, the frequency of the \(i\)-th normal mode, and the symmetry label of that mode.  
For the interstate coupling \(\lambda_i^{(n,m)}\) the feature vector contains the energies of the two electronic states (\(n\) and \(m\)), together with the same vibrational frequency and symmetry.
Having defined the features, we build the dataset matrix \(\mathbf D\) from two sub‑matrices: a feature matrix \(\mathbf X\) and a target vector \(\mathbf Y\).  Each row of \(\mathbf X\) is a feature vector \(\mathbf x_j\); each entry of \(\mathbf Y\) is the corresponding scalar coupling constant (either \(\kappa_i^{(n)}\) or \(\lambda_i^{(n,m)}\)).  If the data set contains \(k\) samples and the feature vector has length \(f\), then

\begin{equation}
\label{eqn:dataset}
\begin{gathered}
  k \in \mathbb{N}, \quad f \in \mathbb{N}, \\[1ex]
  \mathbf{x}_j = (x_{j1}, \dots, x_{jf}) \in \mathbb{R}^{1\times f}, \quad j = 1, \dots, k \\[1ex]
  \mathbf{X} =
    \begin{bmatrix}
      \mathbf{x}_1 \\ \vdots \\ \mathbf{x}_k
    \end{bmatrix}
    \in \mathbb{R}^{k\times f}, \quad
  \mathbf{Y} =
    \begin{bmatrix}
      y_{1} \\ \vdots \\ y_{k}
    \end{bmatrix}
    \in \mathbb{R}^{k\times 1}, \\[1ex]
  \mathbf{D} = 
    \bigl[\;\mathbf{X}\;\bigm|\;\mathbf{Y}\bigr]
    \;\in \mathbb{R}^{k\times (f+1)}
\end{gathered}
\end{equation}

\begin{figure}[htbp]
\centering
\includegraphics[width=0.80\columnwidth]{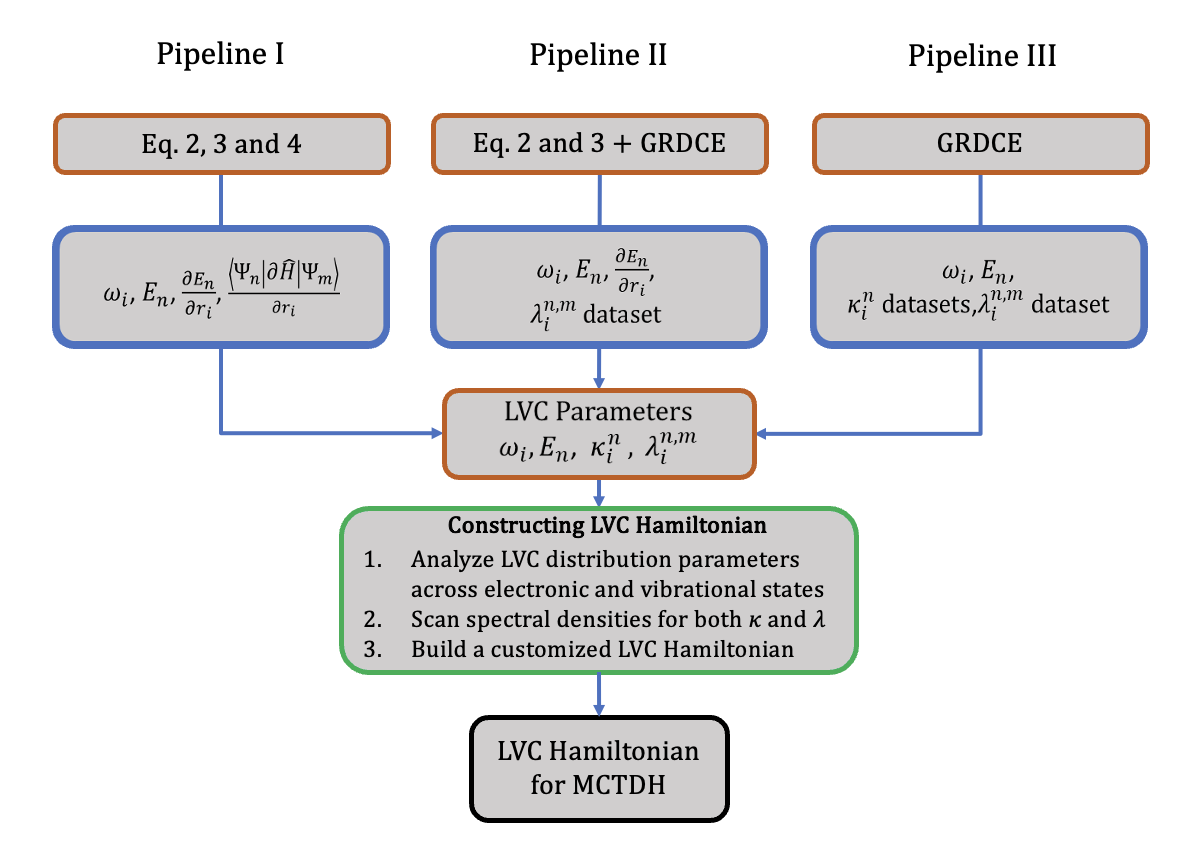}
\caption{PyPC workflow and three pipelines for generating LVC parameters: Pipeline I (full first-principles up to $\sim$20 atoms), Pipeline II (hybrid FP/GRCDE), and Pipeline III (GRCDE for both $\kappa$ and $\lambda$ for large systems).}
\label{fig:fig2}
\end{figure}

For a target molecule the predictor is "conditioned" on the relevant features by first selecting a subset of rows in \(\mathbf D\) that are similar to the molecule of interest.  The conditional distribution of the coupling constant (\(\kappa_i^{(n)}\) or \(\lambda_i^{(n,m)}\)) is then estimated using "univariate kernel density estimation (KDE)" \cite{Rosenblatt1956,Parzen1962}. The mean of this estimated distribution provides the predicted LVC parameter.  
Figure~\ref{fig:fig2} summarizes the complete PyPC workflow and the three computational pipelines that can be used to obtain the parameters \(\kappa_i^{(n)}\) and \(\lambda_i^{(n,m)}\).  In short, the workflow combines FP calculations and GRCDE to generate the essential LVC parameters efficiently. For small systems the procedure relies entirely on FP data; for medium‑sized systems it uses a hybrid FP‑GRCDE approach; and for large systems it can be fully GRCDE‑driven (additional details are provided in the Supporting Information).

\subsection{First-principles excited states data via time-dependent density functional theory}

In order to obtain accurate FP data, time-dependent density functional theory\cite{ferre2016density} (TD-DFT) based on a linear-response approach can be performed for a limited number of atoms. This approach offers important insights into the plasmonic characteristics, including quantum effects. However, the main challenge lies in the rapid increase in computational cost with the size of nanoparticles and the number of desired excited states.\cite{aleotti2021parameterization} Therefore, in order to alleviate the computational cost, for large systems time-dependent density functional theory based on tight binding (TD-DFTB and TD-DFT+TB) is used to calculate the excited states.\cite{ruger2015efficient} In fact, the efficiency and accuracy of TD-DFT+TB for calculating the plasmonic properties of gold and silver nanoparticles have been investigated in our previous work for about 3000 to 30000 electronic states\cite {asadi2020td}. TD-DFT+TB does not rely on a DFTB parametrization and combines the full DFT ground state with time-dependent tight-binding like approximations to reduce the coupling matrix of Casida’s linear response.\cite{ruger2015efficient} Recently, this method has been extended to obtain the gradient of excited states analytically, optimizing the excited state for noble metal nanoclusters with a threefold reduction in cost compared to full TD-DFT gradients\cite {havenridge2023analytical}.

\subsection{Quantum Dynamics Simulations}

Quantum dynamics simulations were performed with the multi-configuration time-dependent Hartree (MCTDH) approach \citep{meyer1990multi,beck2000multiconfiguration,meyer2009multidimensional} in its multilayer \citep{vendrell2011multilayer} generalization as implemented in the Heidelberg package.\citep{worth2000heidelberg, beck2000multiconfiguration} Here the wavefunction ansatz reads
\begin{equation}
\label{eqn:eq5}
\Psi(q_{1},\cdots ,q_{f},t)=\sum_{j_{1}=1}^{n_{1}}\cdots \sum_{j_{f}=1}^{n_{f}}A_{j_{1}\cdots j_{f}}(t)\prod_{\kappa=1}^{f}\Phi_{j_{k}}^{(\kappa)}(q_{\kappa},t)
\end{equation}
where the time-dependent expansion coefficients are denoted as $A_{j_{1}\cdots j_{f}}(t)$, and the time-dependent basis functions, usually referred to as Single Particle Functions (SPFs) as  $\Phi_{j_{k}}^{(\kappa)}(q_{\kappa},t)$. Each SPF is in turn expressed as a linear combinations in a time-dependent basis with time-dependent coefficients
\begin{equation}
\label{eqn:eq6}
\Phi_{j_{k}}^{(\kappa)}(q_{\kappa},t)=\sum_{i_{k}=1}^{N_{k}}c_{i_{k}j_{k}}^{(\kappa)}(t)\chi_{i_{k}}^{(\kappa)}(q_{k})
\end{equation}
providing a discrete variable representation (DVR) grid. The time-evolution of the corresponding equations of motion is determined using the Dirac–Frenkel variational principle \cite{dirac1930principles,frenkel1934wave}. The multi-layer MCTDH (ML-MCTDH) method \citep{vendrell2011multilayer}, is a powerful extension that reduces the computational cost of MCTDH by generalizing the above equations. Essentially the SPFs in Eq. \ref{eqn:eq5} are now expanded using a recursive framework of MCTDH expansions, creating successive layers of time-dependent coefficients, with a set of primitive grid functions in the lowest layer (\ref{eqn:eq6}). In both MCTDH and the ML-MCTDH extension, calculations must be converged with respect to the number of SPFs to ensure accurate results. In order to facilitate this, the ML-spawning algorithm \citep{mendive2017towards,mendive2020regularizing} provides a systematic approach for dynamically expanding the number of SPFs in each layer on-the-fly.

\subsection{Computational Details}

In this study,
symmetric tetrahedral ($T_d$) clusters 
containing 10, 20, 56, and 120 atoms, 
as well as icosahedral ($I_h$) structures
with 55 and 147 atoms, 
%
were chosen for analysis
since their properties 
have been extensively investigated both theoretically and experimentally.\cite{peng2010reversing,kuisma2015localized,cortie2011synthesis,lópez2013effect} 
The geometrical structures 
were taken from the optimized structures reported in our previous study \cite{asadi2020td}.
For the small $Ag_{20}$ cluster, both the totally symmetric ($T_d$) and lower-symmetry ($C_{3v}$, $C_{3}$, and $C_s$) structures were investigated. Geometry optimizations of the lower-symmetry structures were initiated from configurations previously reported in the literature.\cite{mckee2017density,schira2019localized}

\textbf{First principles calculations}. 
Ground-state optimized structures and Hessian matrices were computed using density functional theory (DFT) within the generalized gradient approximation (GGA) and the BP86 exchange–correlation functional \cite{perdew1986density}. 
Depending on cluster size 
excited-state properties were evaluated either using Casida’s TD-DFT formalism \cite{casida1995time} or the free-parameter TD-DFT+TB approximation \cite{ruger2016tight},  which reduces the cost of TD-DFT gradient evaluations by roughly a factor of three. 
DFT and TD-DFT calculations were carried out using the ORCA 5.0\cite{neese2012orca,neese2018software} and Q-Chem\cite{epifanovsky2021software} packages, while TD-DFT+TB calculations were performed using the AMS 2023 package.\cite{ADF2001}

For the small $Ag_{10}$ cluster, excited-state energies, gradients, and non-adiabatic interstate couplings (NACs)\cite{fatehi2011analytic,ou2015first} were computed using the Casida TD-DFT method and the hybrid PBE0 xc-functional\cite{adamo1999toward} and the def2-TZVPP basis set. \cite{schafer1992fully,weigend2005balanced} TD-DFT was employed to calculate vertical excited-state energies and transition dipole moments for the medium-sized clusters ($Ag_{20}$, $Ag_{55}$, and $Ag_{56}$) and TD-DFT+TB for the larger  ($Ag_{120}$ and $Ag_{147}$) clusters.
These computations employed triple-ζ polarized (TZP or TZ2P) Slater-type basis sets\cite{van2003optimized} with the pure PBE \cite{perdew1996generalized} or hybrid PBE0 \cite{adamo1999toward} xc-functionals, implemented within the scalar relativistic ZORA formalism.\cite{van1996zero,lenthe1993relativistic}
Remarkably, these clusters exhibit $10^2 - 10^3$ electronic states in the energy range of approximately 1.0 to 5.0 eV.
For large plasmonic nanoclusters, we have previously shown that the TD-DFT+TB approach \cite{ruger2015efficient, ruger2016tight} delivers nearly 100-fold greater efficiency than linear-response TD-DFT while maintaining electronic accuracy within 0.15 eV \cite{asadi2020td}. Moreover, the excited-state gradients can be computed within TD-DFT+TB 
through an analytical formulation as implemented by Havenridge et al. in the AMS 2023 package\cite{havenridge2023analytical,ADF2001}. 
The accuracy of TD-DFT+TB in calculating excited-state gradients relative to full TD-DFT is assessed in the Supporting Information by comparing the absolute values of $\kappa_i^{(n)}$. For the $a_{1}$ and $e$ modes of Ag$_{20}$, the mean absolute values predicted by TD-DFT+TB (PBE) are roughly twice those obtained with full TD-DFT (PBE0).

\textbf{Quantum dynamics simulations}. Quantum dynamics propagations were run using the ML-MCTDH method as implemented in the Heidelberg MCTDH package.\cite{worth2000heidelberg,beck2000multiconfiguration} A harmonic oscillator DVR was employed for all degrees of freedom, with sizes in between 15 to 30 primitive basis functions. 
ML-MCTDH calculations must be converged with respect to the number of SPFs and the size of the primitive basis to ensure accurate results. 
This was verified by ensuring that the lowest natural orbital populations remained below $10^{-6}$,
and that the grid point populations were sufficiently $10^{-9}$. 
%
In addition, the observables of interest, such as expectation values of the coordinates, diabatic populations and the shape of the absorption spectra were examined to converge respect to the size of the wavefunction. In all propagations the wavepacket was propagated for 400 fs. and absorption spectra were obtained from the Fourier transform of the corresponding autocorrelation function, $\left\langle \Psi(t)|\Psi(0) \right\rangle$.
Finally, an analysis of efficiency and scalability of the overall protocol as function of size of the system is detailed in Table S1 and Figure S2 and S3 of the Supporting Information.   

\section{Results and Discussion}
This section is divided into three parts. First, we describe the extraction of LVC parameters from ground- and excited-state calculations. We analyse the distribution of these coefficients with respect to the size of the clusters and discuss how the pipelines proposed in Figure 2 can be used to either: i) derive a lower-dimensionality Hamiltonian that still preserves the original distributions of the LVC parameters, or ii) extrapolate a Hamiltonian for larger clusters based on the distributions of parameters for smaller clusters. 
For example, the symmetry of vibrational modes is one of the important features that is not always easily obtained from FP data, but that instead can be estimated from a smaller data set. A second important point is that the distribution of first-order LVC parameters turns out to be independent to the electronic state, thus facilitating their sampling. Following this, in the next subsection, the effectiveness of the  protocol is demonstrated by successfully reproducing the experimental absorption spectra \cite{schira2018effects,harb2008optical} and vibronic broadening of the medium and large plasmonic Ag$_{20}$, Ag$_{55}$ and Ag$_{120}$ clusters. The respective population dynamics of plasmonic states are then investigated, showing that a LVC model derived as a statistical distribution from an appropriate dataset accurately predicts ultrafast lifetimes for bright states and effectively captures the dynamics of dark states. 

\subsection{The Parameters of Model Hamiltonian}
\begin{figure}[ht!]
    \centering
    \begin{subfigure}[t]{0.45\textwidth}
        \centering
        \includegraphics[width=\textwidth]{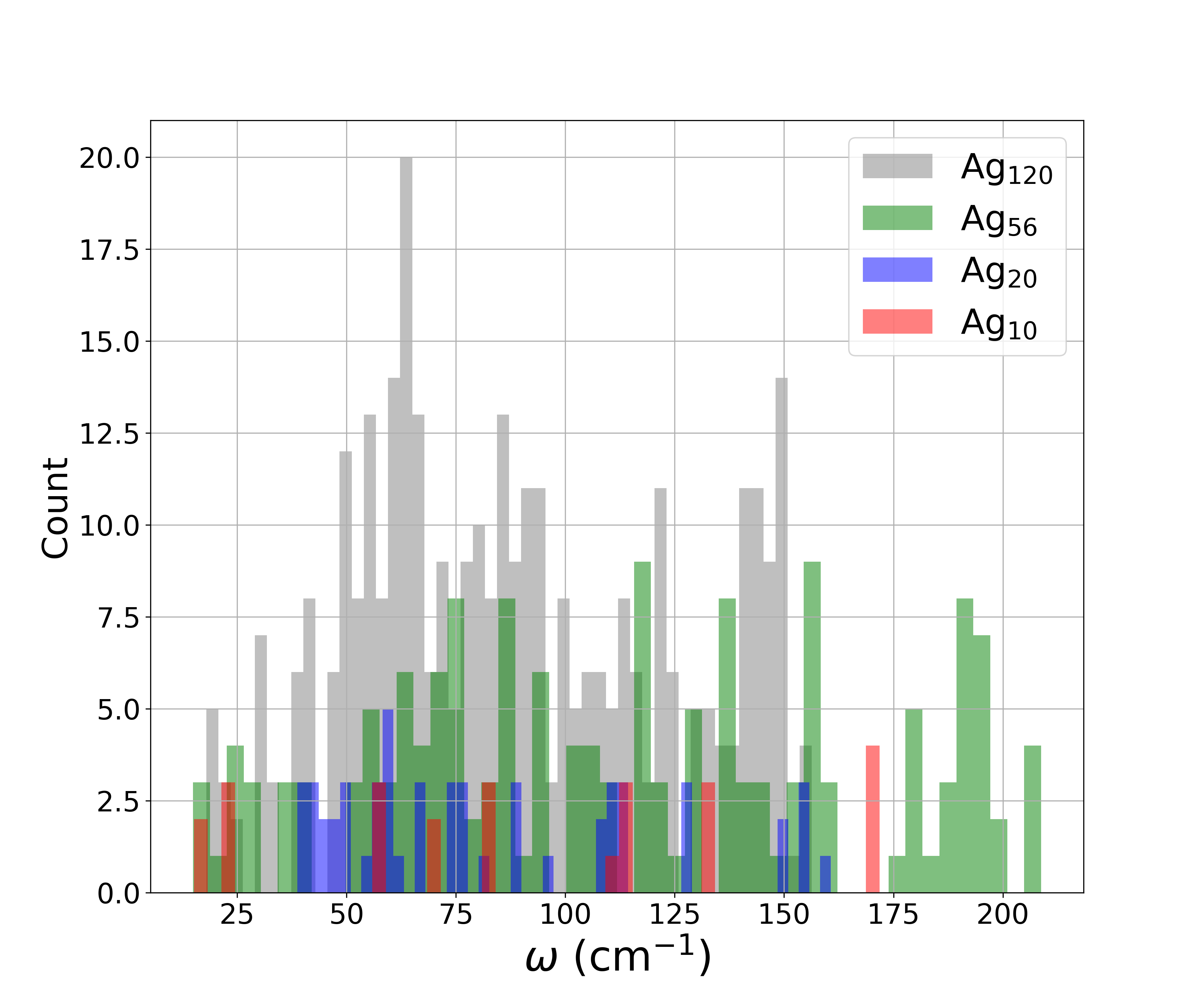}
        \caption{}
        \label{fig:fig3a}
    \end{subfigure}
    \vspace{1.0em} 
    \begin{subfigure}[t]{0.45\textwidth}
        \centering
        \includegraphics[width=\textwidth]{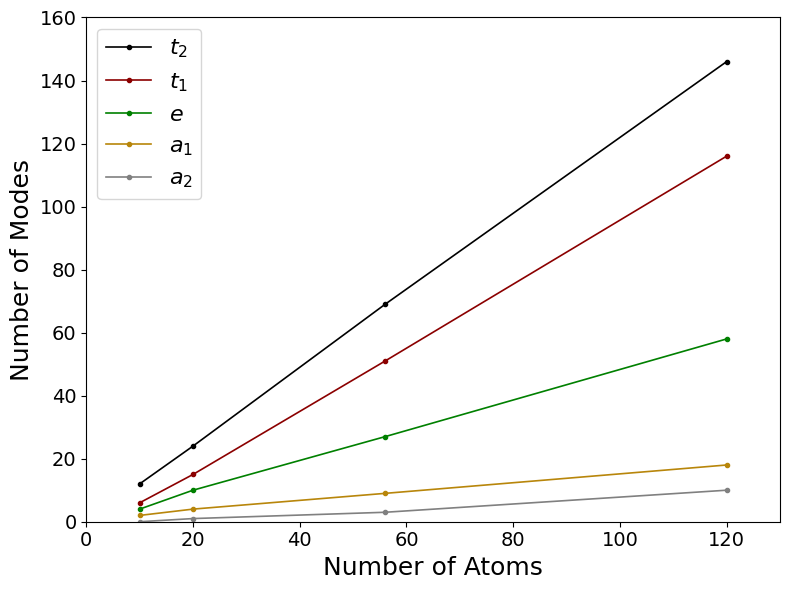}
        \caption{}
        \label{fig:fig3b}
    \end{subfigure}
    \vspace{1.0em} 
    \begin{subfigure}[t]{0.45\textwidth}
        \centering
        \includegraphics[width=\textwidth]{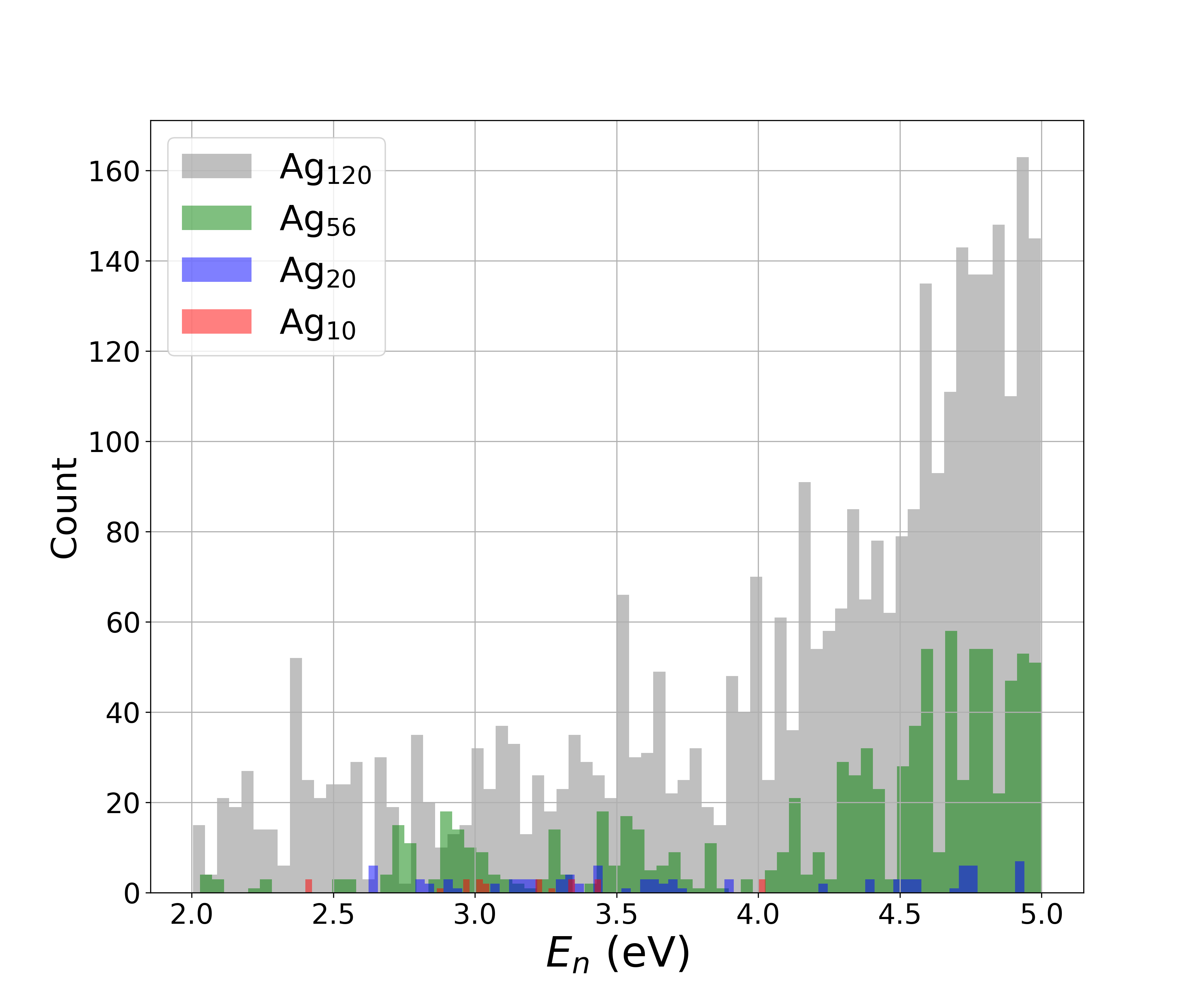}
        \caption{}
        \label{fig:fig3c}
    \end{subfigure}
    \vspace{1.0em}
    \begin{subfigure}[t]{0.45\textwidth}
        \centering
        \includegraphics[width=\textwidth]{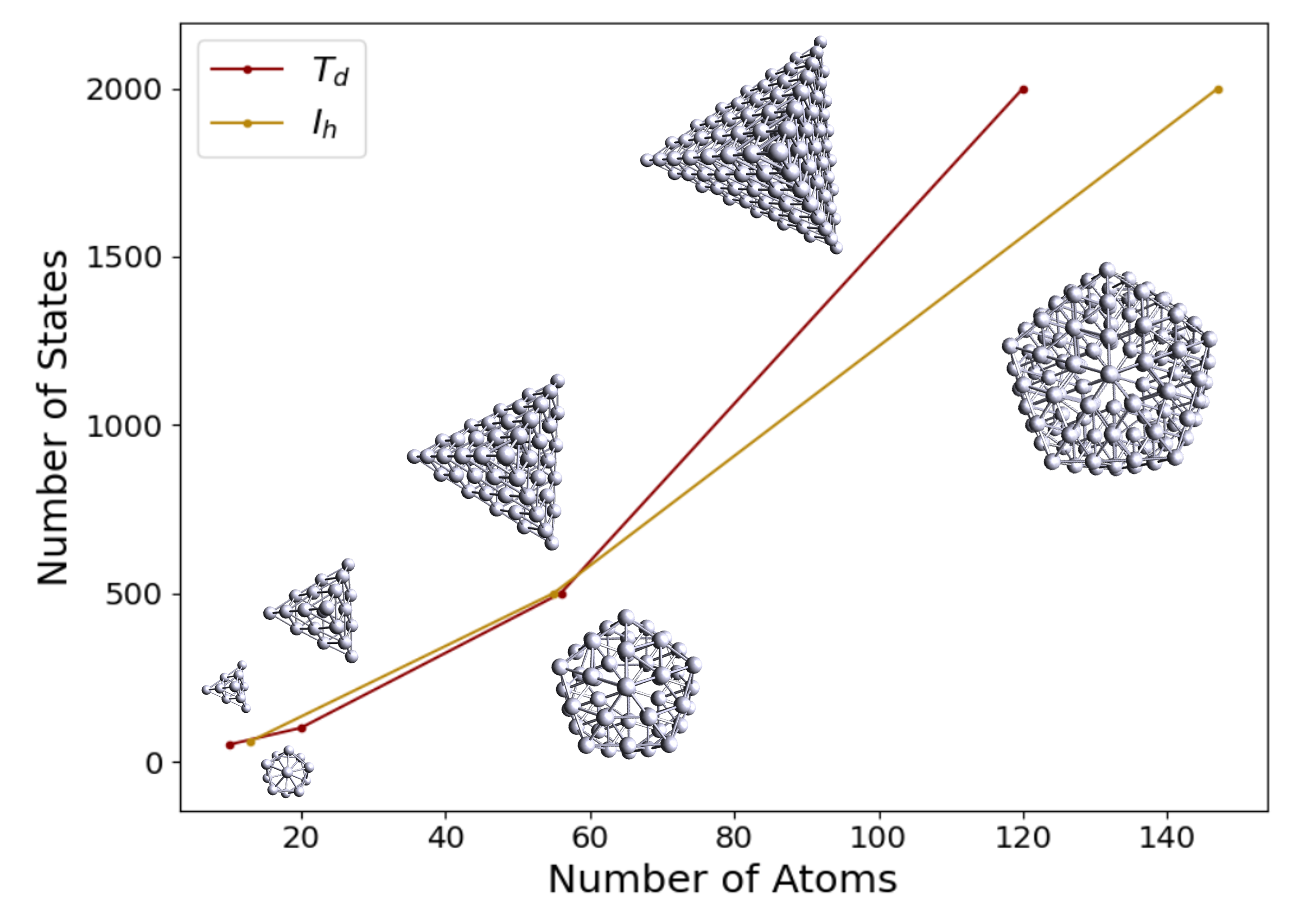}
        \caption{}
        \label{fig:fig3d}
    \end{subfigure}
    \caption{For tetrahedral Ag$_{10}$, Ag$_{20}$, Ag$_{56}$, and Ag$_{120}$ clusters: (a) Histogram of the vibrational frequency distribution, $\omega$; 
    (b) Linear relationship between the number of vibrational modes and the number of atoms, categorized by irreducible representations: t$_{2}$, t$_{1}$, e, a$_{2}$, and a$_{1}$; (c) Histogram of the excited state energy distribution, $E_{n}$; 
    (d) Variation in the number of collective plasmonic excited states with nanoparticle size for tetrahedral and icosahedral symmetries at the level of TD-DFT+TB.}
    \label{fig:figure3}
\end{figure}

\textit{Zeroth-order LVC parameters} --- The reference system, i.e. the zeroth-order Hamiltonian, is defined by the harmonic vibrational frequencies, $\omega _{i}$, and the energetic origin of each diabatic electronic state, $E_{n}$ at the chosen reference geometry.
In the following we analyze the corresponding distributions of these parameters for tetrahedral clusters Ag$_{n}$ ($n=$ 10, 20, 56 and 120).
Figure \ref{fig:fig3a} shows the histogram distribution of
vibrational modes $\omega _{i}$
in the range of 20 to 200 cm$^{-1}$; the corresponding numerical values are detailed in Tables S2–S6 of the Supporting Information.
We focus on this range since far infrared spectra from 100 to 250 cm$^{-1}$ have been reported for size-selected silver clusters using free-electron laser radiation.\cite{van2017structural}
As we can see, the smallest clusters Ag$_{10}$, and Ag$_{20}$ show discrete spikes at irregular spacing. However, analysis of the distribution of vibrational modes for lower-symmetry ($C_{3v}$, $C_{3}$, and $C_{s}$) Ag$_{20}$ clusters shows 
that as the symmetry decreases, for example from $T_{d}$ to $C_{s}$, the modes become more uniformly distributed in the 20–160 cm$^{-1}$ range (see Figure S4b in the Supporting Information).
Similarly, as the cluster increases in size, the distribution becomes smoother, broad an continuous 
across most of the range.
This is expected because as the cluster grows, atoms in the interior begin to approximate the bulk lattice environment, while only surface atoms remain under-coordinated. 
Thus, in principle, the histogram of frequencies tends toward the phonon density of states of bulk of silver (face-centered cubic, fcc).

Importantly, this convergence with increasing cluster size towards a well-defined limit
enables the estimation of wavenumber values for large-sized clusters based on Hessian calculations performed for smaller-sized clusters.
In this regard, Figure \ref{fig:fig3b} shows the linear scaling between the number of vibrational modes in each irreducible representation of the $T_{d}$ point group ($a_{1}$, $a_{2}$, $e$, $t_{1}$, $t_{2}$) and the number of atoms in Ag$_{n}$ clusters ($n=$ 10, 20, 56 and 120).
It should be noted that there is a remarkable linear correlation.
Recognizing the symmetry of each vibrational mode is essential for correctly extracting the first-order terms of the LVC. 
However, as cluster size increases, assigning symmetry to individual modes becomes increasingly difficult, specially for Hessian computations in large basis sets. 
Therefore, this pattern facilitates not only the prediction of the total number of modes for larger nanoparticles as a whole, but also for each irreducible representation and the corresponding proportional ratios when devising reduced-dimensionality Hamiltonians.

On the other hand, Figure \ref{fig:fig3c} shows the density of excited states between 2.0–5.0 eV at the TD-DFT+TB level for the same clusters. 
As we can see, qualitatively the number of states in this range increases rapidly with cluster size.
In contrast with the previous case, the electronic density of states does not have a upper bound since electrons can be excited to arbitrarily high energies.
This is confirmed in Figure \ref{fig:fig3d} by plotting the number of states as a function of system size, which in this case is not linear,
requiring roughly 100, 500, and 2000 states for $Ag_{20}$, $Ag_{56}$, and $Ag_{120}$, respectively.
Naturally, the electronic basis can be restricted to those electronic states
exhibiting plasmonic character as in our previous studies.\cite{asadi2020td}
Gas-phase absorption spectra of small- and medium-sized $Ag_{n}$ clusters ($n=$ 5–147) reveal strong plasmonic excitations between 3.8–4.1 eV, which redshift slightly with increasing size \cite{schira2019localized,yu2018optical,harb2008optical}. 
The experimental linewidth (FWHM) ranges from 0.3–0.5 eV, with narrower widths observed in clusters of high-symmetry geometries.
Therefore, the number of states
in the electronic basis 
required to
capture the plasmonic excitation window (3.7–4.5 eV) can be much smaller. For instance, the required number is dramatically reduced by including roughly 15, 60, and 500 states for $Ag_{20}$, $Ag_{56}$, and $Ag_{120}$, respectively.

\begin{figure}[H]
 \centering
 \resizebox{0.8\textwidth}{!}{
  \begin{minipage}{\textwidth}
    \begin{subfigure}[t]{0.45\textwidth}
      \includegraphics[width=\textwidth]{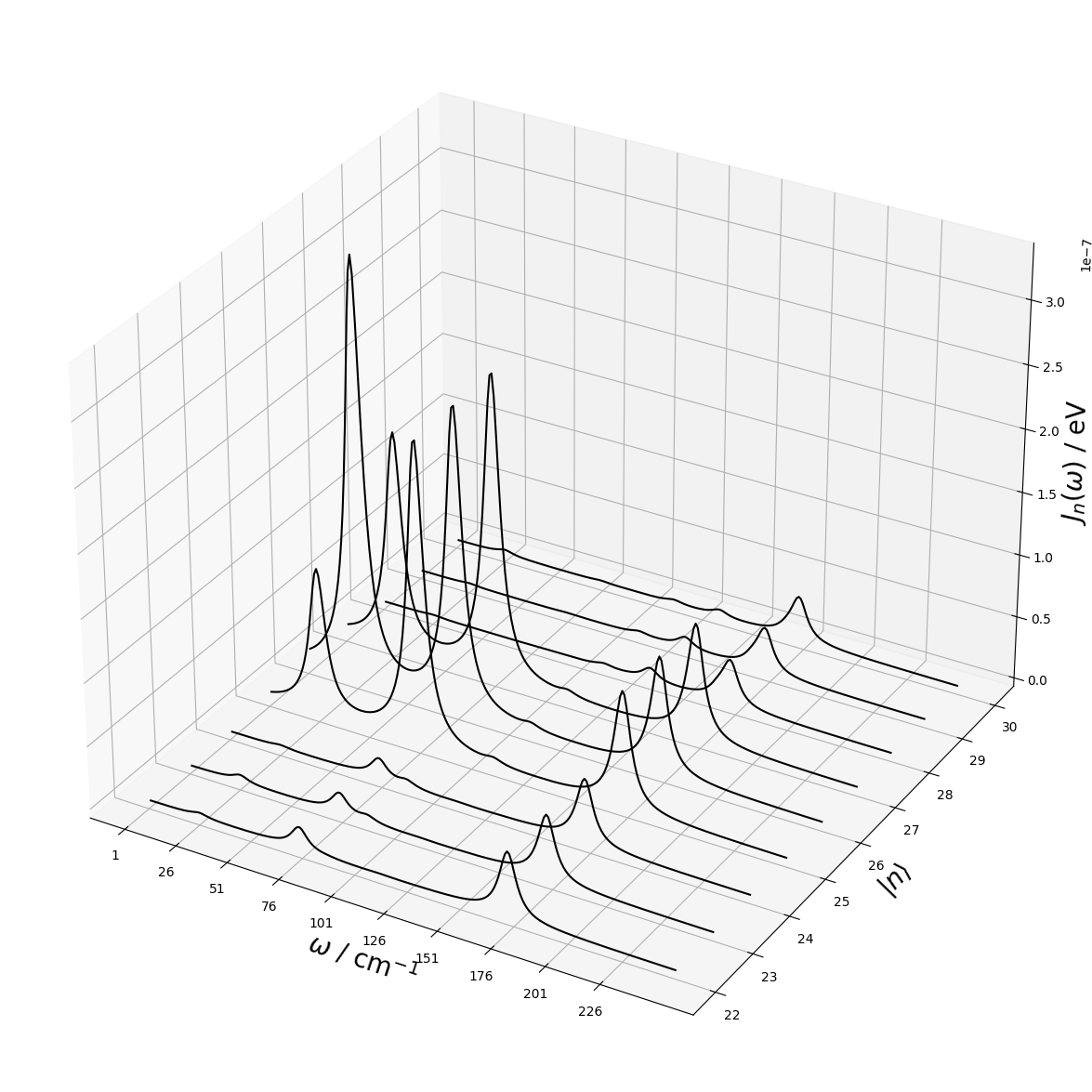}
      \caption{}
      \label{fig:fig4a}
    \end{subfigure}
    \begin{subfigure}[t]{0.45\textwidth}
      \includegraphics[width=\textwidth]{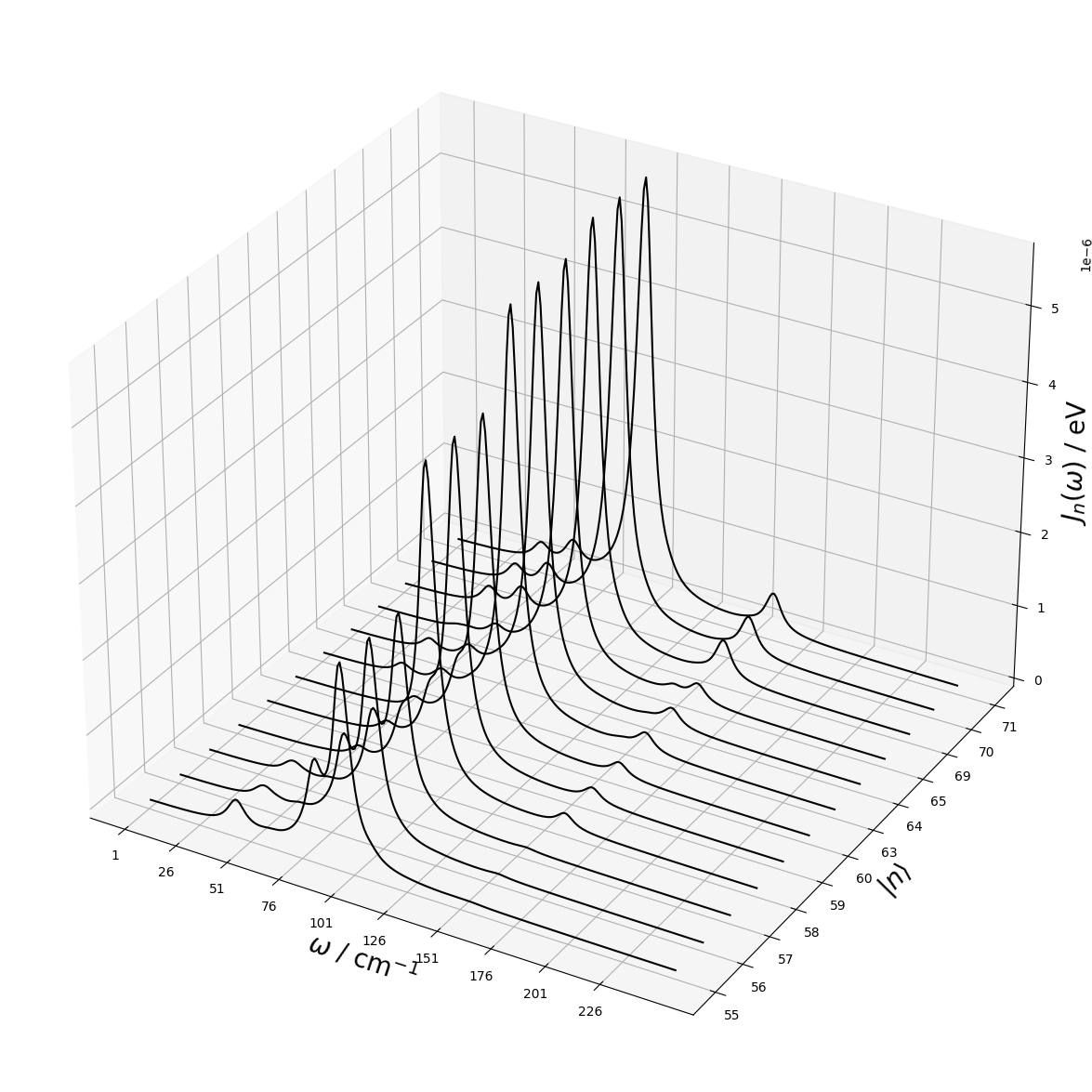}
      \caption{}
      \label{fig:fig4b}
    \end{subfigure}
    
    \begin{subfigure}[t]{0.45\textwidth}
      \includegraphics[width=\textwidth]{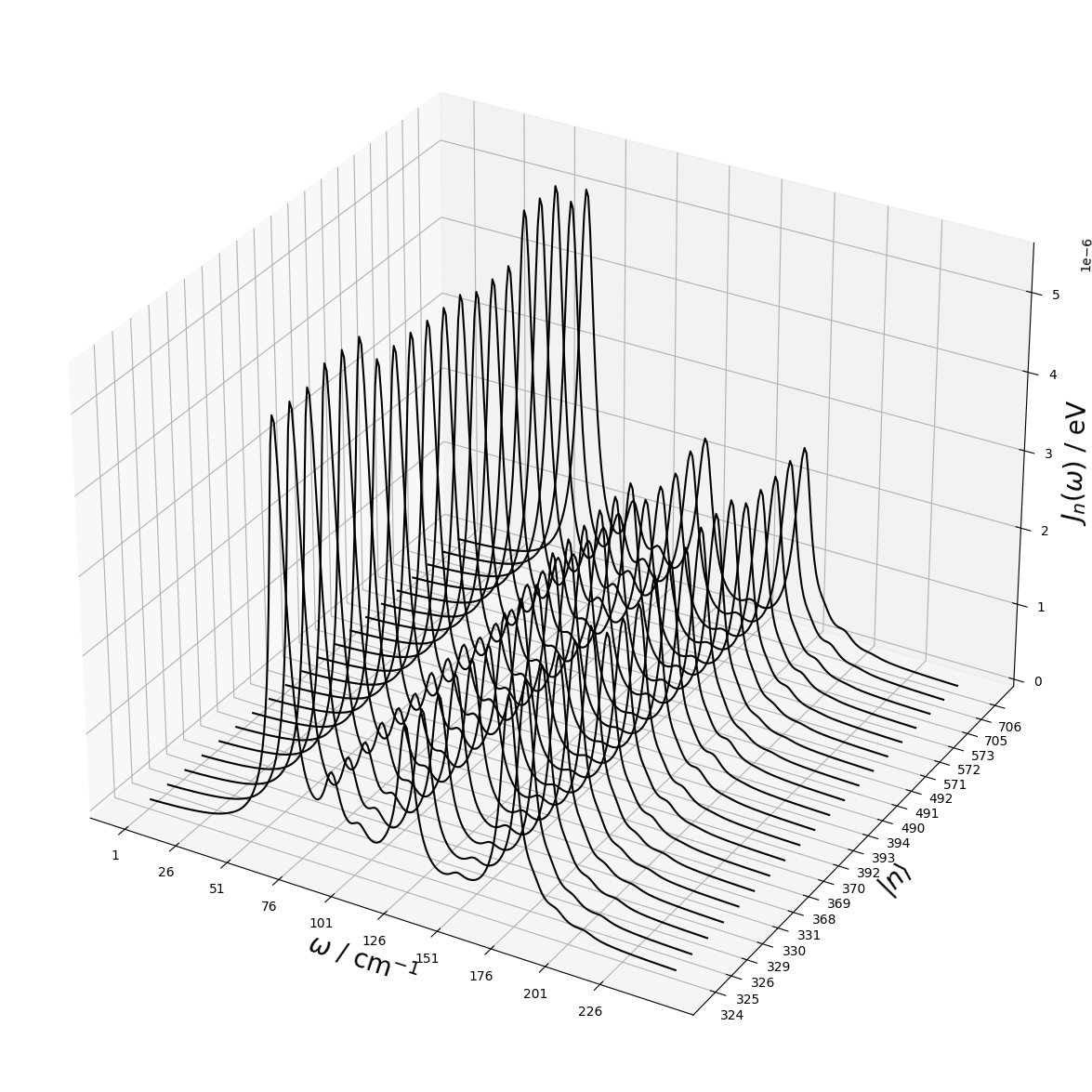}
      \caption{}
      \label{fig:fig4c}
    \end{subfigure}
    \begin{subfigure}[t]{0.42\textwidth}
      \includegraphics[width=\textwidth]{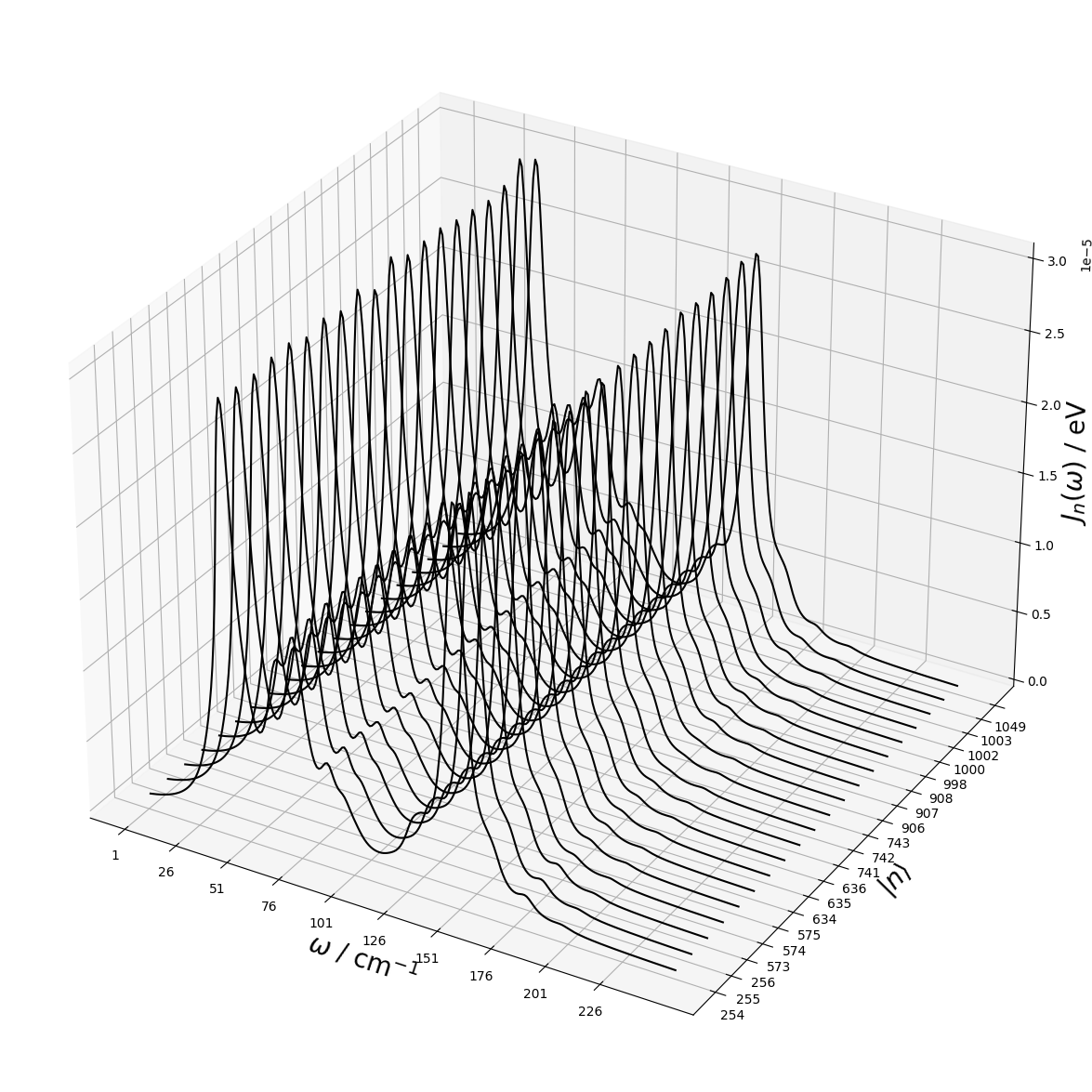}
      \caption{}
      \label{fig:fig4d}
    \end{subfigure}
    
    \begin{subfigure}[t]{0.45\textwidth}
     \includegraphics[width=\textwidth]{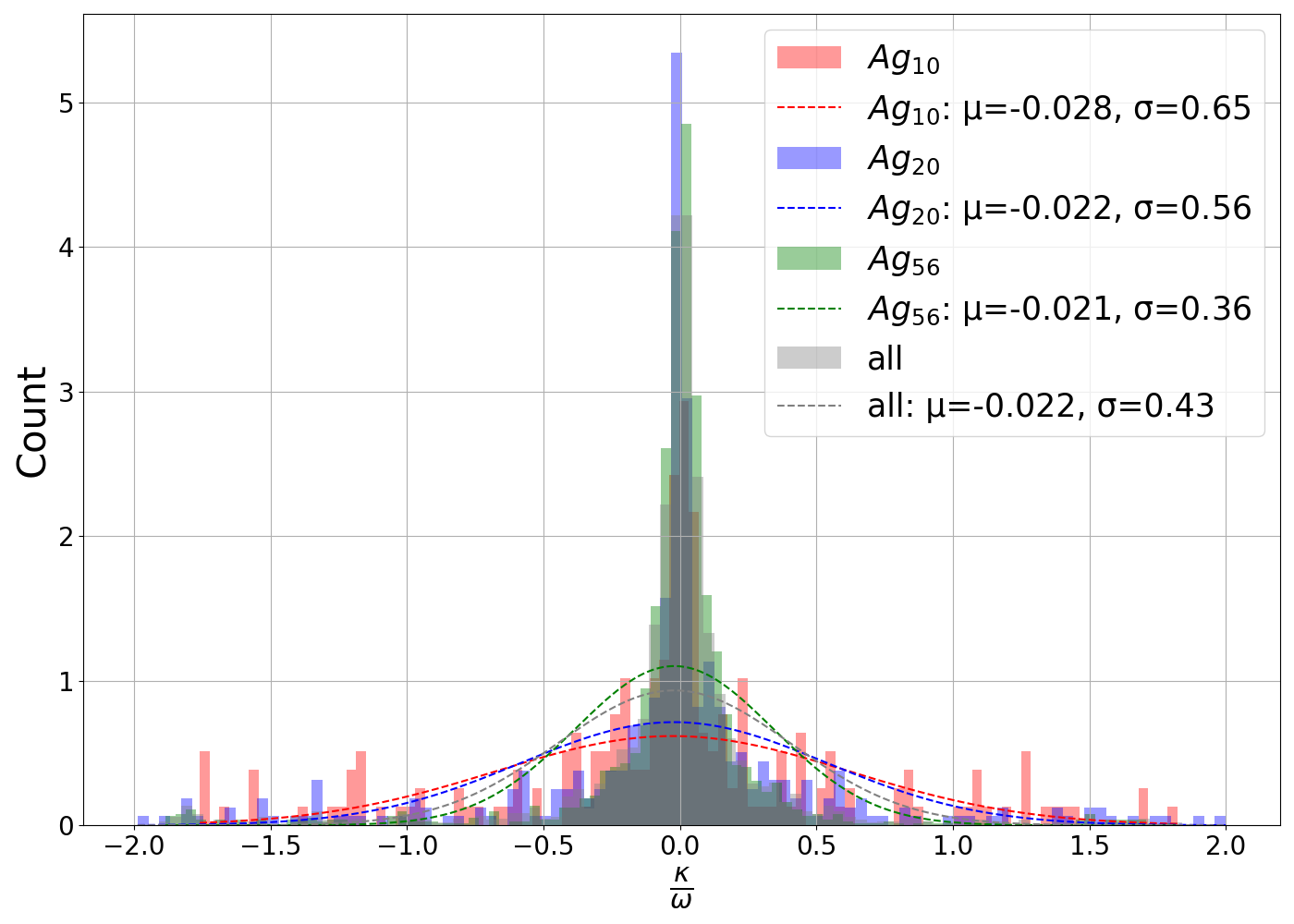}
     \caption{}
     \label{fig:fig4e}
    \end{subfigure}
    \begin{subfigure}[t]{0.45\textwidth}
      \includegraphics[width=\textwidth]{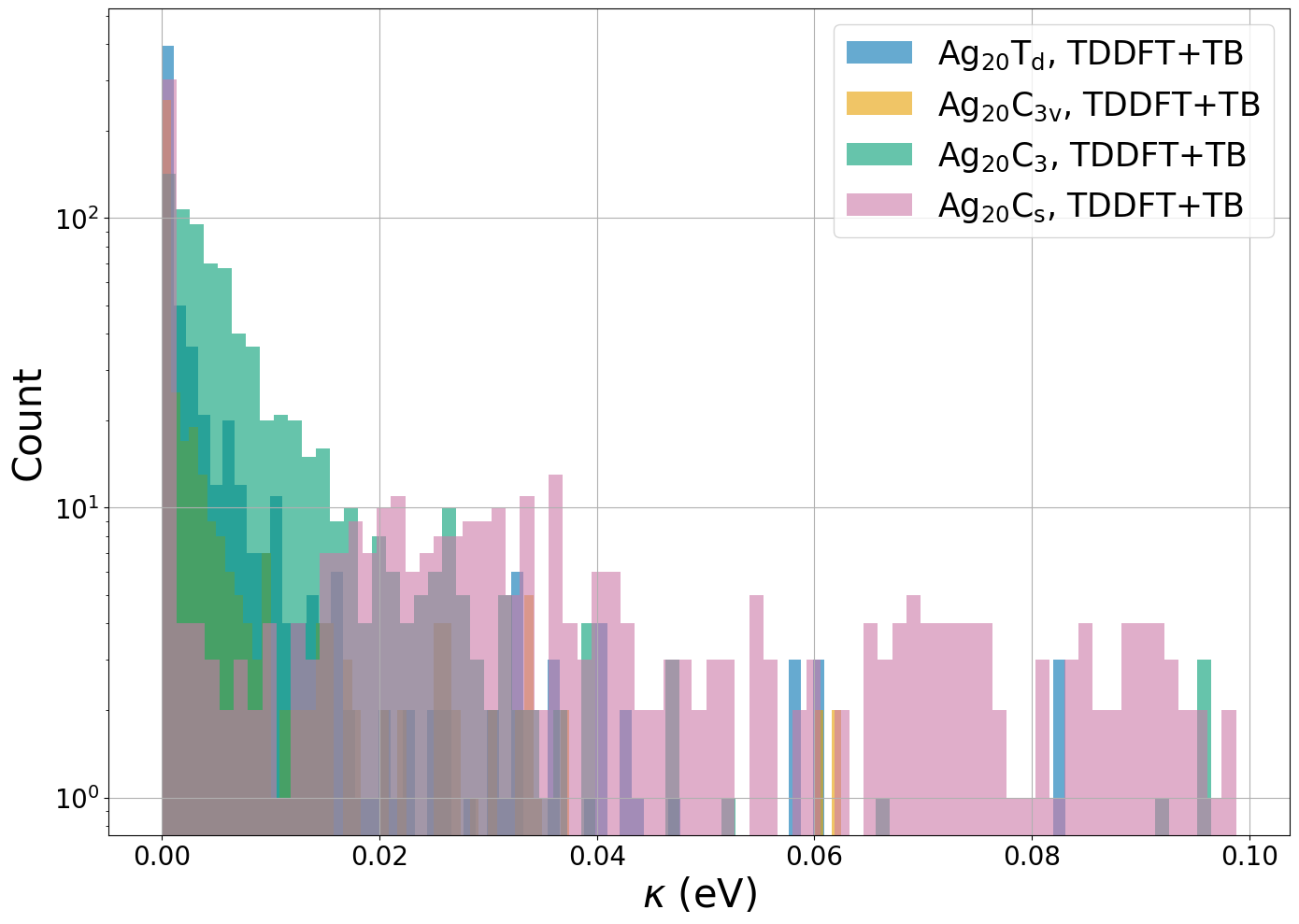}
      \caption{}
      \label{fig:fig4f}
    \end{subfigure}
    \label{fig:fig4}
  \end{minipage}
   }
    \caption{Variation of the spectra density of $\kappa_{i}^{(n)}$ for increasingly large clusters, i.e. a) $Ag_{10}$, b) $Ag_{20}$, c) $Ag_{56}$ (tetrahedral) and d) $Ag_{55}$ (icosahedral). (e) Gaussian distribution of $\frac{\kappa}{\omega}=\frac{\kappa_{i}^{(n)}}{{\omega_{i}}}$ at the TD-DFT+TB level of theory. (f) Histogram distribution of $\kappa=|\kappa_{i}^{(n)}|$ (in $eV$), for $Ag_{20}$ with $T_{d}$, $C_{3V}$, $C_{3}$ and $C_{s}$ symmetries, at TD-DFT+TB level of theory.}
    \label{fig:figure4}
\end{figure}

\textit{First-order on-diagonal LVC parameters}---
In the following we discuss the on-diagonal $\kappa_i^{(n)}$ coefficients, which in contrast with the previous case depend on two descriptors, both a vibrational mode $i$ and an electronic state $n$. Thus, a convenient quantity to analyze this bi-dimensional distribution is the so-called "spectral density", $J_n(\omega)$,\cite{Green2024CNB} defined as:
\begin{equation}
\label{eqn:eq7}
J_{n}(\omega)=\frac{\pi}{2}\sum_{i=1}^{3N-6}\left( \kappa_{i}^{n} \right)^{2}\delta(\omega-\omega_{i})
\end{equation}

which is a function that encodes how strongly the vibrational mode of frequency $\omega_i$ couple to the electronic state $n$.
Figures \ref{fig:fig4a}, \ref{fig:fig4b} and \ref{fig:fig4c} 
present $J_n(\omega)$ within the plasmonic excitation range for the tetrahedral clusters $Ag_{10}$, $Ag_{20}$ and $Ag_{56}$, respectively.
A small cluster such as $Ag_{10}$
presents significant variation across different electronic states.
However, as the size of the cluster increases, once again we see that
$J_n(\omega)$ displays much less variation
for relatively large clusters such as $Ag_{20}$ and $Ag_{56}$.
Moreover, the same tendency is observed for the icosahedral cluster $Ag_{55}$ (in Figure \ref{fig:fig4d}), thus supporting that, this is a tendency present in large clusters, independently of the symmetry point group.  
Consequently, justifying the treatment of the on-diagonal $\kappa_i^{(n)}$ terms as a one-dimensional distribution that is independent of the specific plasmonic electronic state. 
This distribution can then be used to generate linear coupling values for the vibrational modes sampled in the previous section. 

The total number of gradient coefficients is dictated by the product of the number of electronic states and nuclear vibrational modes considered.
However, due to symmetry considerations, these will be non-zero only if the symmetry of the vibrational mode corresponds to the totally symmetric irreducible representation of the molecule's point group. 
For example, for the $Ag_{20}$ and $Ag_{56}$ clusters, which have 54 and 162 vibrational modes, and at least 15 and 40 selected electronic states (within the plasmonic band range), the number of $\kappa_i^{(n)}$ parameters is 810 and $6,480$, respectively. This number can then increase rapidly, reaching approximately $28,320$ terms for $Ag_{120}$, which has 354 vibrational modes, and for which we select 80 electronic states at the PBE0/TZP level of theory.

Nonetheless, a much smaller number of parameters can be generated stochastically, while still preserving the original distributions of $\kappa_i^{(n)}$ and $\omega_{i}$.
A distribution of gradient coefficients for the bright ($T_{2}$) and dark ($T_{1}$, $E$) electronic states, coupled with $a_{1}$, $e$, and $t_{2}$ vibrational modes, was generated from all the $Ag_{10}$, $Ag_{20}$, and $Ag_{56}$ available data. 
%
%
This dataset was normalized by calculating the ratio $\kappa_{i}^{(n)}/\omega_{i}$, as shown in Figure \ref{fig:fig4e}. A histogram of $\kappa_{i}^{(n)}/\omega_{i}$ was then constructed and fitted with a normal distribution, yielding to a mean ($\mu$) and standard deviation ($\sigma$) characterizing each cluster and the whole dataset comprising over 5,000 values. 

Last, in order to have more insight into the role of symmetry over these distributions, 
Figure \ref{fig:fig4f} compares the distribution of $\left| \kappa_{i}^{n} \right|$ for Ag$_{20}$ in its totally symmetric form ($T_{d}$) to its lower symmetry configurations ($C_{3V}$, $C_{3}$, and $C_{s}$) at the TD-DFT+TB level of theory.
As the symmetry decreases from $T_{d}$ to $C_{s}$, the number of vibrational modes with $\left| \kappa_i^{(n)} \right| \geq 0.01$ eV increases markedly, from 4 modes of $a_{1}$ symmetry to 54 modes of $a'$ and $a''$ symmetries. In lower-symmetry clusters, $\left| \kappa_i^{(n)} \right|$ values are nearly uniformly distributed between 0.01–0.1 eV. This uniformity in medium-sized clusters, thus
facilitates 
estimating $\kappa_{i}^{(n)}/\omega_{i}$ ratios in larger, low-symmetry systems.
The more evenly spaced is the parent distribution, i.e. uniform, the more stable and faster convergence will be present in the estimates.

\clearpage 

\textit{First-order off-diagonal LVC parameter} -- 
For $Ag_{10}$, we considered the interstate couplings between 9 electronic states (with $T_{2}$ and $T_{1}$ symmetries) and $3N-6$ vibrational modes, resulting in a total of 864 $\lambda_i^{(n,m)}$ terms.
For the coupling of bright states with $T_{2}$ symmetry, the product is $T_{2}\otimes T_{2}=A_{1}\oplus E\oplus T_{2}$, resulting in non-zero interstate coupling terms for vibrational modes with $a_{1}$, $e$ and $t_{2}$ symmetries.
The selected electronic dark states (DS) of $Ag_{10}$ possess $T_{1}$ symmetry. Consequently, the BD ($T_{2}\otimes T_{1}$) and DD ($T_{1}\otimes T_{1}$) interstate couplings are non-zero for vibrational modes with $a_{1}$, $e$, $t_{1}$ and $t_{2}$ symmetries.
The symmetry and wavenumber of vibrational modes with the largest interstate couplings, $\left| \lambda_i^{(n,m)} \right|$, are listed in Supporting Information (Figure S5 and Table S12.)

\begin{figure}[ht!]
    \centering
    \begin{subfigure}[t]{0.32\textwidth}
        \centering
        \includegraphics[width=\textwidth]{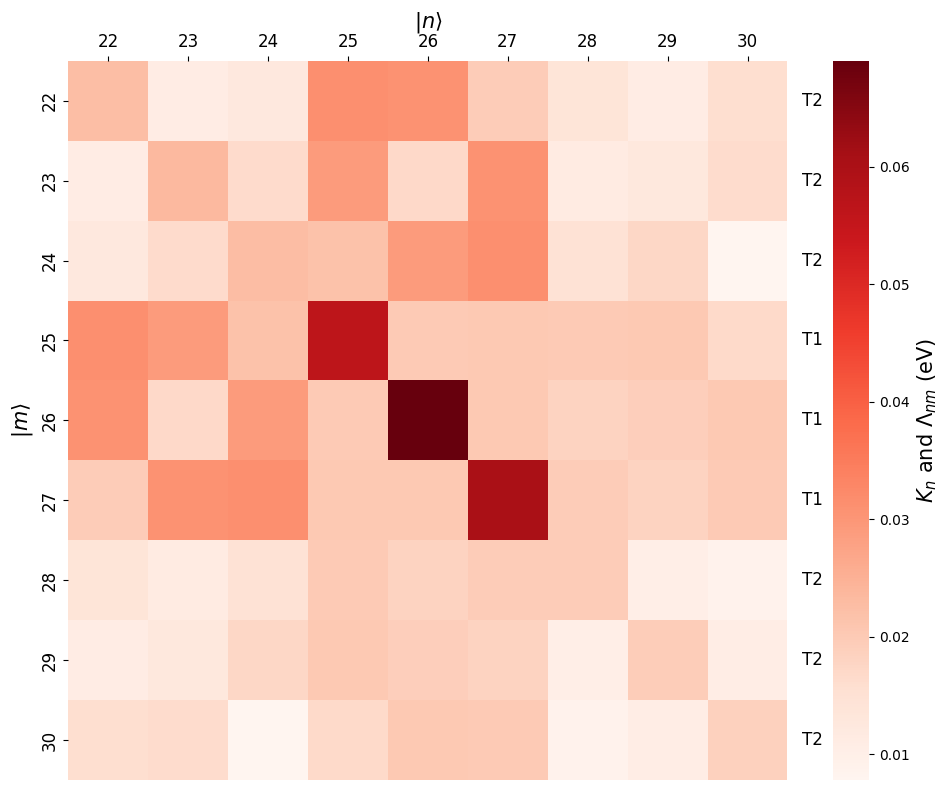}
        \caption{}
        \label{fig:fig5a}
    \end{subfigure}
    \vspace{1.0em} 
    \begin{subfigure}[t]{0.32\textwidth}
        \centering
        \includegraphics[width=\textwidth]{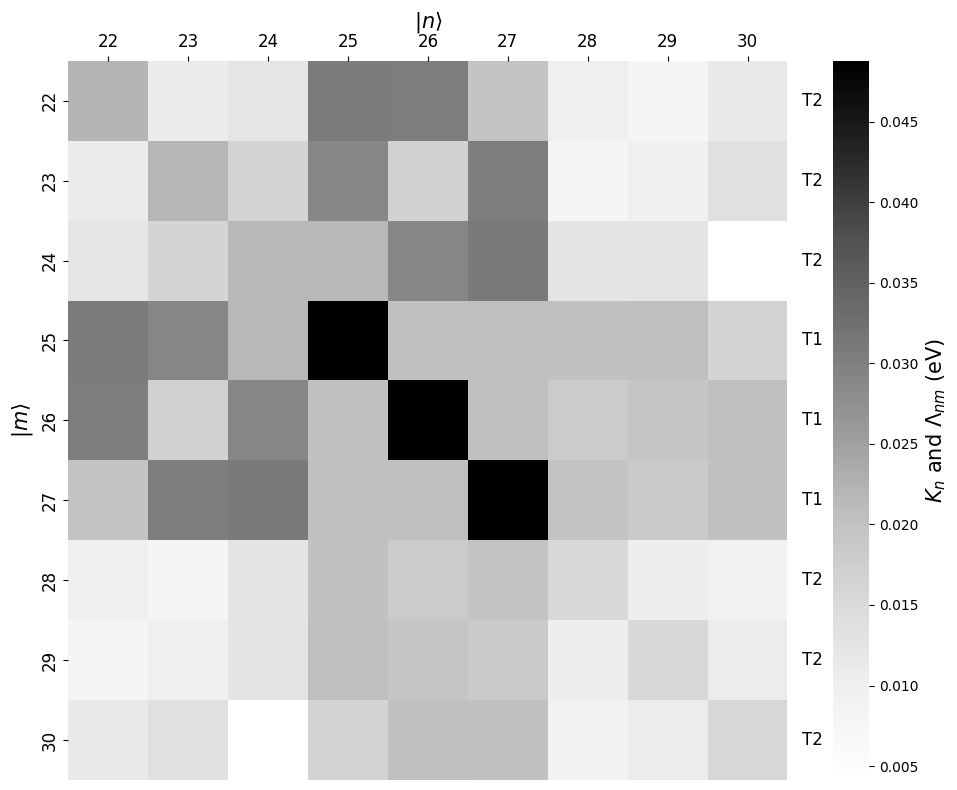}
        \caption{}
        \label{fig:fig5b}
         \end{subfigure}
     \vspace{1.0em} 
    \begin{subfigure}[t]{0.32\textwidth}
        \centering
        \includegraphics[width=\textwidth]{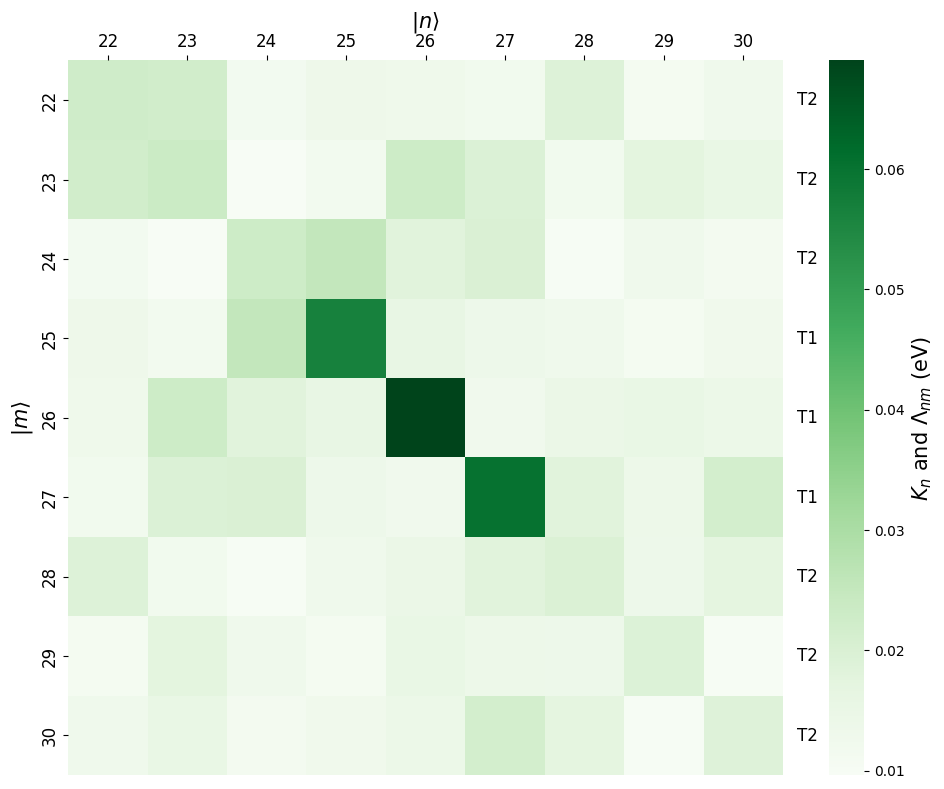}
        \caption{}
        \label{fig:fig5c}    
    \end{subfigure}
   \caption{Vibronic coupling maps illustrating the effective couplings for diagonal entries ($K_{n}$) and off-diagonal entries ($\Lambda_{nm}$) for electronic states with $T_{1}$ and $T_{2}$ symmetries. Panel (a) Full set of intrastate coupling ($\kappa$) and interstate coupling terms ($\lambda$) from FP. (b) Reduced set of $\kappa$ and $\lambda$ terms from FP. (c) Reduced set of $\kappa$ from FP and $\lambda$ from statistical distribution on dataset, pipeline ~II.}
    \label{fig:figure5}
\end{figure}

For $Ag_{10}$, the effective vibronic couplings—$K_{n} = \left[ \sum_{i=1}^{I} \left( \kappa_{i}^{n} \right)^{2} \right]^{1/2}$ for diagonal terms and $\Lambda_{nm} = \left[ \sum_{i=1}^{I} \left( \lambda_{i}^{nm} \right)^{2} \right]^{1/2}$ for off-diagonal terms \cite{Green2024CNB}—are constructed by combining contributions from individual vibrational modes into collective modes and their corresponding couplings. Figure \ref{fig:fig5a} and \ref{fig:fig5b} show the matrix representation of the vibronic coupling maps for the bright ($T_{2}$) and dark ($T_{1}$) states extracted from the pipeline~I. The difference between Figure \ref{fig:fig5a} and \ref{fig:fig5b} is in the number of vibrational DOFs and coupling parameters. In Figure \ref{fig:fig5a} all $\kappa_{all}^{(n)}= 216$ and $\lambda_{all}^{(n,m)}= 864$ terms are included, and in Figure \ref{fig:fig5b} the symmetry constraints and imposing the threshold on the amount of values are applied and the number of coupling parameters decreases to $\kappa_{a_{1}, e}^{(n)} = 54$ and $\lambda_{t_{2}, e}^{(n,m)} = 576$.
In addition, Figure \ref{fig:fig5c} displays the matrix representation of the vibronic coupling map when the generated $\lambda(e, t_{2})$ values are from pipeline~II.
The diagonal couplings among dark states ($S_{25}$–$S_{27}$, $T_{1}$ symmetry) are stronger than those between bright states ($S_{22}$–$S_{24}$ and $S_{28}$–$S_{30}$, $T_{2}$ symmetry).
For off-diagonal term the mixed-symmetry states (i.e., between dark and bright states), the coupling values of dark–bright fall between the dark–dark and bright–bright terms, consistent with values in Figure S5. These analyses provide a systematic approach to identifying the descriptors for both intra- and interstate coupling terms.

\subsection{Quantum Dynamics and Absorption Spectra}
\begin{figure}[ht!]
    \centering
    \begin{subfigure}[t]{0.30\textwidth}
        \centering
        \includegraphics[width=\textwidth]{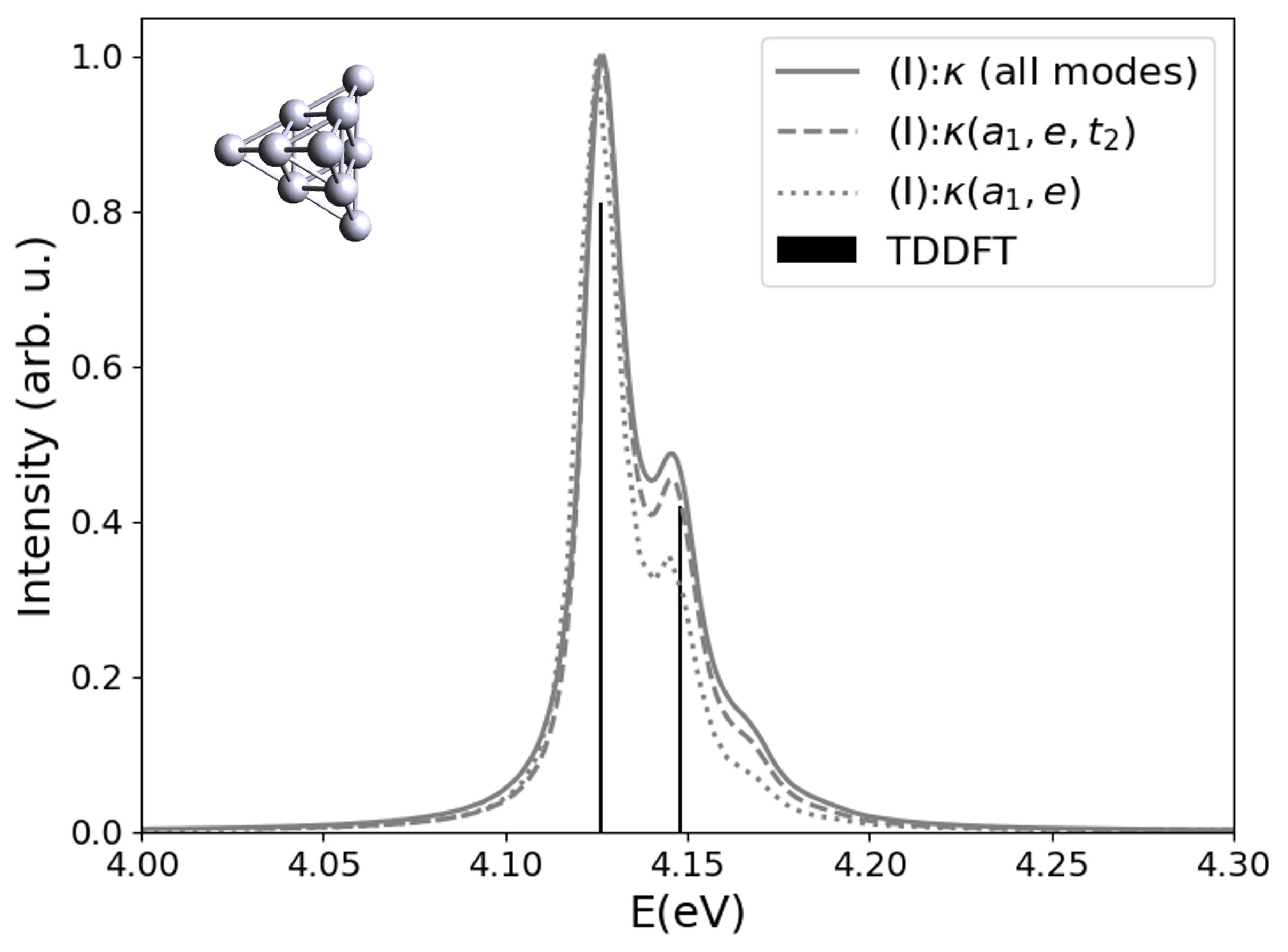}
        \caption{}
        \label{fig:fig6a}
    \end{subfigure}
    \vspace{1.0em} 
    \begin{subfigure}[t]{0.30\textwidth}
        \centering
        \includegraphics[width=\textwidth]{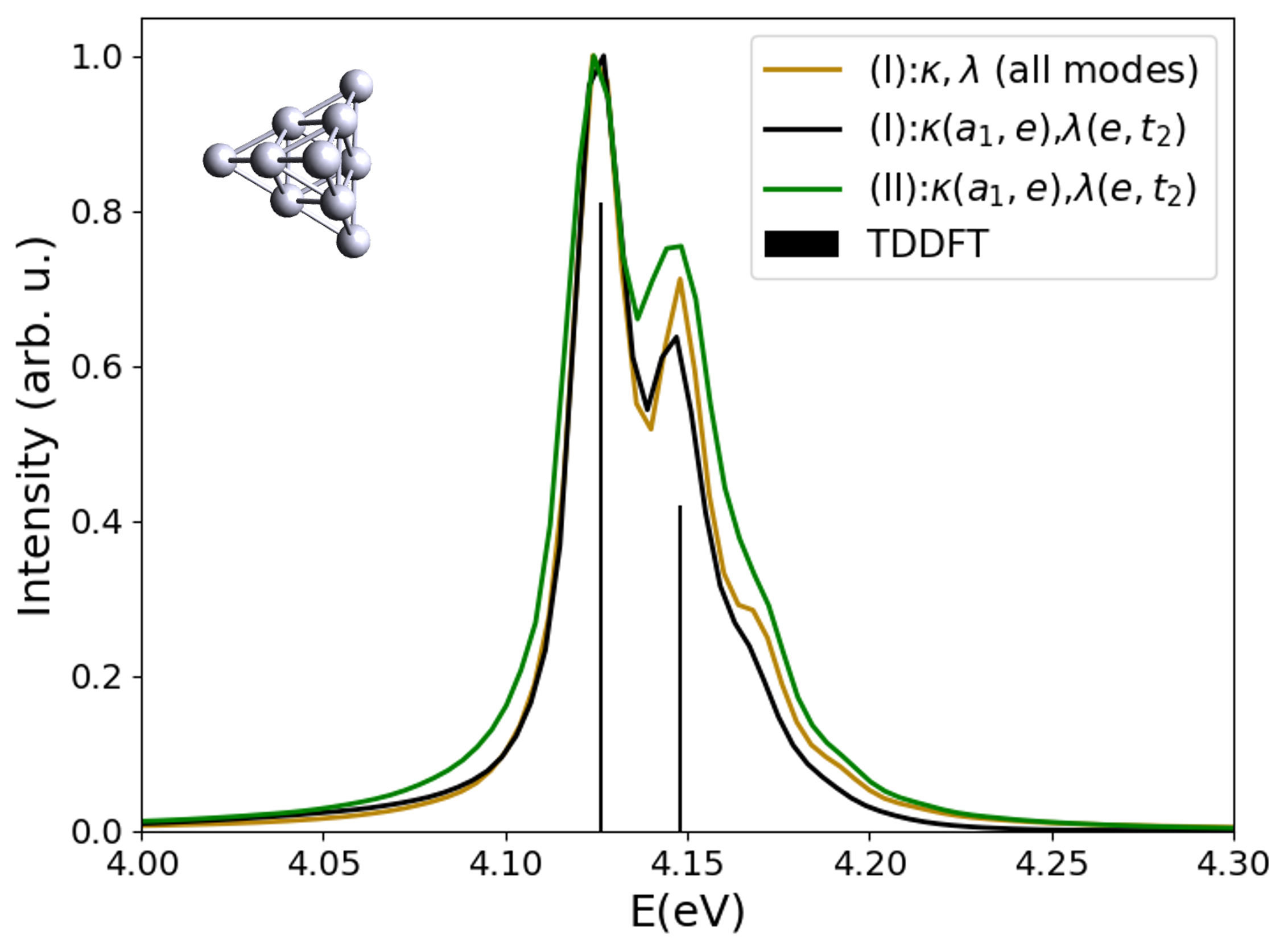}
        \caption{}
        \label{fig:fig6b}
         \end{subfigure}
     \vspace{1.0em} 
    \begin{subfigure}[t]{0.30\textwidth}
        \centering
        \includegraphics[width=\textwidth]{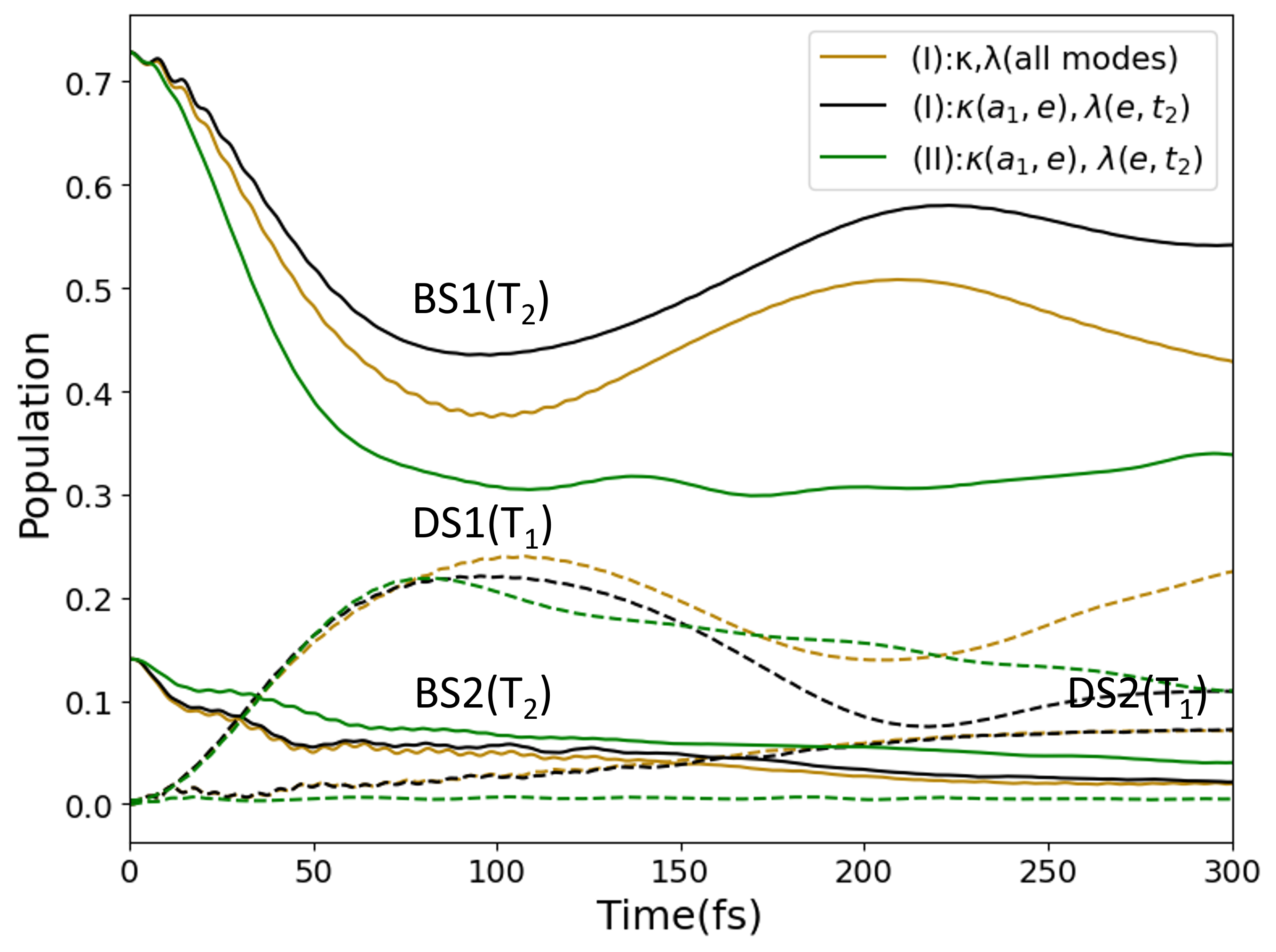}
        \caption{}
        \label{fig:fig6c}    
    \end{subfigure}
   \caption{Absorption spectra and population dynamics of $Ag_{10}$. Panel (a) illustrates the broadening of the spectrum when intrastate coupling ($\kappa$) is considered for a reduced number of modes compared to the full set. Panels (b) and (c) show the effects of the interstate coupling term ($\lambda$) on the absorption spectrum and population dynamics, respectively. The ($\lambda$) values obtained from non-adiabatic coupling calculations are shown for all modes (yellow line), for $e$ and $t_{2}$ modes (black line), and are compared with ($\lambda$) derived from the statistical distribution, $\lambda (e, t_{2})$(II) (green line).
   For these spectra the damping function $f(t)=e^{-t/\tau}$, with the damping time  $\tau=150$ fs was applied.}
    \label{fig:figure6}
\end{figure}

\textit{Benchmark calculations for small $Ag_{10}$ cluster}--
As outlined in the previous section, the number of parameters in the vibronic Hamiltonian is determined by the number of vibrational modes and electronic states. Notably, the efficiency of quantum dynamics calculations is highly sensitive to the total number of vibrational modes involved.
In this part, we aim to assess the effect of parameter reduction by imposing a threshold on $\left| \lambda_i^{(n,m)} \right|$ and limiting the number of vibrational modes based on symmetry considerations. The analysis focuses on the impact of these reductions on spectral patterns and population dynamics.
For the Ag$_{10}$ cluster with $T_d$ symmetry, quantum dynamics calculations were performed using ML-MCTDH on nine electronic states and 24 vibrational modes. 
The wavepacket was initially excited (vertically to the excited states) and propagated over 400 fs. The absorption spectra, derived from the Fourier transform of the autocorrelation function, are shown in Figure \ref{fig:fig6a} and \ref{fig:fig6b}.

The broadening of the absorption spectrum originates from the intrastate coupling terms, whereas the TD‑DFT calculation yields a discrete “stick” spectrum (Figure \ref{fig:fig6a}). For a LVC model that includes nine electronic states and 24 vibrational modes, the full set of intrastate coupling parameters comprises 216 terms (solid line). To examine the effect of symmetry selection we also show spectra obtained with reduced coupling sets: a set of 162 terms that contain only the  ${a_{1},t_{2},}$ and ${e}$ symmetries (dashed line) and a further reduced set of 54 terms limited to the $a_{1}$ and ${e}$ symmetries (dotted line).

Figure \ref{fig:fig6a} displays the contributions of vibrational modes with $a_{1}$, $e$ and $t_{2}$ symmetries to intrastate broadening.
Notably, modes with $a_{1}$ and $e$ symmetries primarily contribute to the broadening of the spectrum and the $t_{2}$ modes have the negligible contribution to the broadening of peaks.
The role of interstate coupling in shaping the spectral broadening and population dynamics is shown in Figures \ref{fig:fig6b} and \ref{fig:fig6c}. 
When all 216 intrastate coupling constants $\kappa_i^{(n)}$ and the 864 interstate coupling constants $\lambda_i^{(n,m)}$ are initially retained, the absorption spectrum is obtained from the full Hamiltonian (yellow line). After imposing the symmetry constraints – i.e. keeping only the $\kappa_i^{(n)}$ terms with $a_1$ and $e$ symmetry (54 terms) and the $\lambda_i^{(n,m)}$ terms with $t_2$ and $e$ symmetry (576 terms) -- the spectrum generated with the reduced Hamiltonian (black line) reproduces the essential features of the full-Hamiltonian spectrum.
This indicates that the essential physics of the vibronic couplings can be captured with a symmetry-reduced parameter set, substantially lowering computational cost without sacrificing accuracy. Importantly, when the generated values of $\lambda(e, t_{2})$ from pipeline~II ((II)-$\lambda(e, t_{2})$, green line) are compared with the FP absorption spectrum ((I)-$\lambda(e, t_{2})$, black line), the spectral broadening remains in close agreement with the exact calculation. This demonstrates that the machine-learned parameters can faithfully reproduce the effect of interstate couplings on the absorption spectrum, validating the predictive power of the PyPC framework.

To further evaluate the lifetime of excited states, Figure \ref{fig:fig6c} shows the electronic population dynamics of four representative states—two bright states (BS1, BS2) and two dark states (DS1, DS2)—over a propagation time of 300 fs. For the bright states, BS1 and BS2 , the population transfer profiles are nearly identical when comparing the full LVC Hamiltonian with its symmetry-reduced counterpart, confirming that symmetry constraints preserve the essential dynamics. Notably, BS1 undergoes a rapid loss of population accompanied by a corresponding gain in DS1, whereas BS2 exhibits a slower depletion that is mirrored by a gradual increase in DS2. This contrast highlights different pathways of population relaxation depending on the bright–dark state coupling. 
To assess the predictive accuracy of generated parameters, we compared population dynamics obtained using interstate couplings from pipeline~II ((II)-$\lambda(e, t_{2})$, green line) with those from FP parameters ((I)-$\lambda(e, t_{2})$, black line). 
The results show satisfactory agreement within the first 100 fs, where most of the ultrafast relaxation occurs, and beyond 100 fs, the populations remain nearly constant. 
This indicates that the generated parameters faithfully capture the ultrafast dynamics, validating the reliability of the PyPC framework for modeling population transfer in plasmonic nanoclusters.

\textit{Absorption spectra for medium and large clusters}--
LVC models are a promising approach for obtaining the vibronic spectra of moderately sized clusters, provided that the electronic structure calculations required to parameterize the Hamiltonian are not a bottleneck. However, as the size of metal clusters and the number of vibrational modes increase, efficiently and accurately calculating the non-adiabatic terms, which needed to parameterize the off-diagonal elements of the LVC model, becomes increasingly computational demanding.
The diagonal terms of the LVC Hamiltonian can be efficiently evaluated for Ag$_{20}$, Ag$_{56}$ and Ag$_{55}$ clusters using analytical excited-state gradient calculations with the TD-DFT+TB method.

The experimental UV absorption spectrum of neutral Ag$_{20}$ is dominated by a broad peak spanning 3.5–4.1 eV, consistent with the characteristic plasmonic response of small silver clusters. \cite{yu2018optical}
Notably, experimental observations suggest that non-totally symmetric geometries are often more stable than the idealized high-symmetry $T_d$ structure. 
To capture this effect in our model Hamiltonian, we explicitly evaluated the intrastate coupling parameters, $\kappa_i^{(n)}$, for Ag$_{20}$ clusters with reduced symmetries ($C_{3v}$ and $C_{s}$). 
These calculations were performed using both TD-DFT (PBE0) and TD-DFT+TB (PBE) level of calculations, enabling a systematic comparison between high- and low-symmetry configurations as well as across different theoretical levels (see Figure S6a and S6b).
As illustrated in Figure \ref{fig:fig4f}, lowering the symmetry of Ag$_{20}$ clusters markedly increases the number of vibrational modes with non-zero $\kappa_i^{(n)}$, rising from only 14 modes in the $T_{d}$ structure to 54 modes in the $C_{s}$ configuration (see Table S15 in Supporting Information). 
This expansion in the set of active modes directly contributes to the broadening of the absorption spectra, as evidenced by the comparison between high- and low-symmetry structures in Figure S6c. 
The results highlight the crucial role of reduced symmetry in enhancing vibronic coupling and reproducing the experimentally observed spectral width.
These spectra were obtained without including the interstate coupling terms, $\lambda_i^{(n,m)}$. 

To faithfully reproduce the shape of the experimental absorption spectrum and accurately capture the population decay dynamics, it is essential to include the interstate coupling terms. These couplings play a pivotal role in determining the dissipative character of plasmonic cavities by mediating non-radiative decay pathways. As the cluster size increases, the electronic manifold becomes increasingly dense, with many states clustered within a narrow energy window (see Figure \ref{fig:fig3c}). This dense manifold of electronic states is strongly coupled to vibrational modes, which are primarily distributed within the low-frequency range of 25–250 cm$^{-1}$ (see Figure \ref{fig:fig3a}), thereby facilitating efficient vibronic energy redistribution.
%

\begin{figure}[ht!]
    \centering
    \begin{subfigure}[t]{0.45\textwidth}
        \centering
        \includegraphics[width=\textwidth]{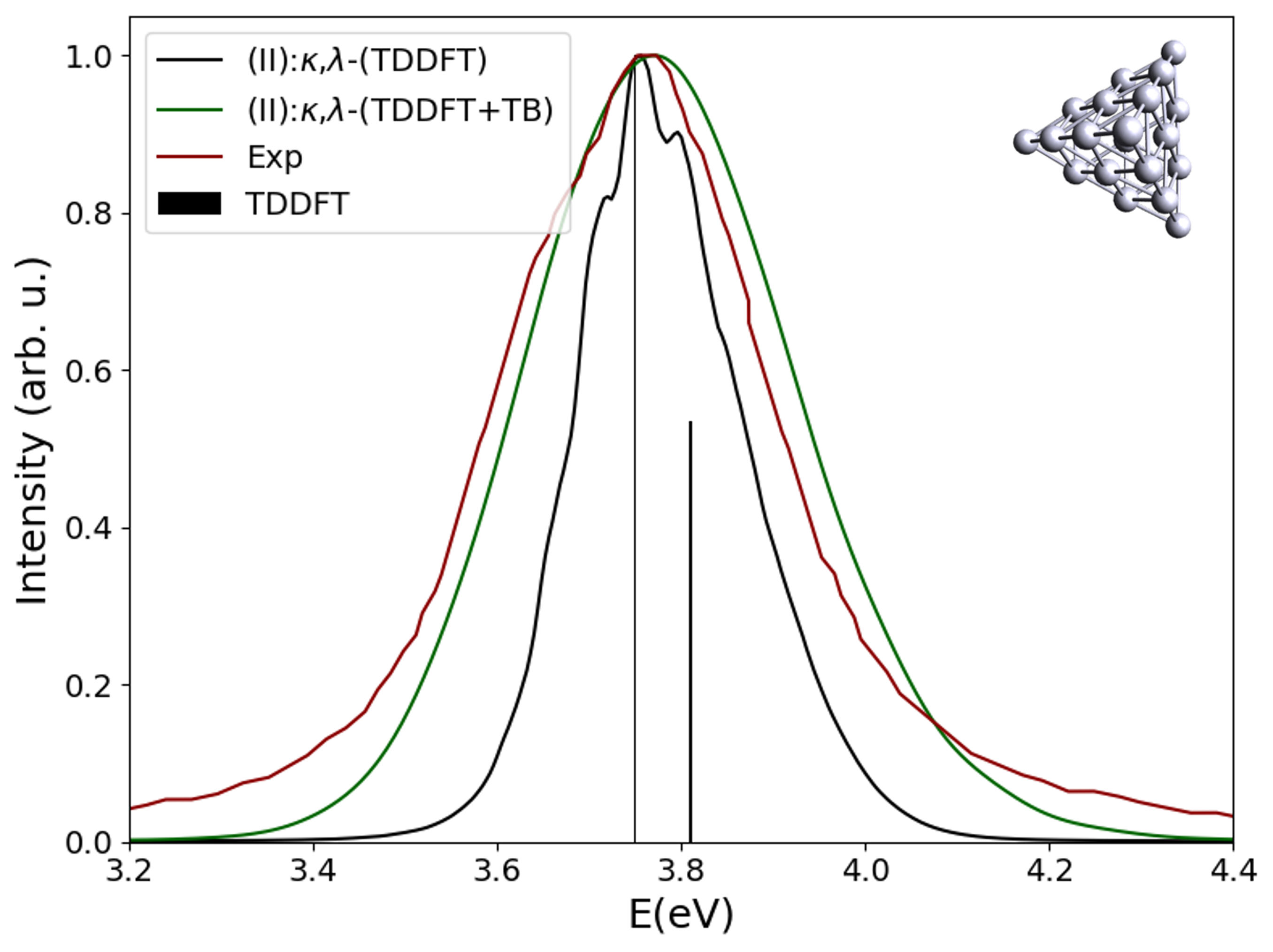}
        \caption{}
        \label{fig:fig7a}
    \end{subfigure}
    \vspace{1.0em} 
    \begin{subfigure}[t]{0.45\textwidth}
        \centering
        \includegraphics[width=\textwidth]{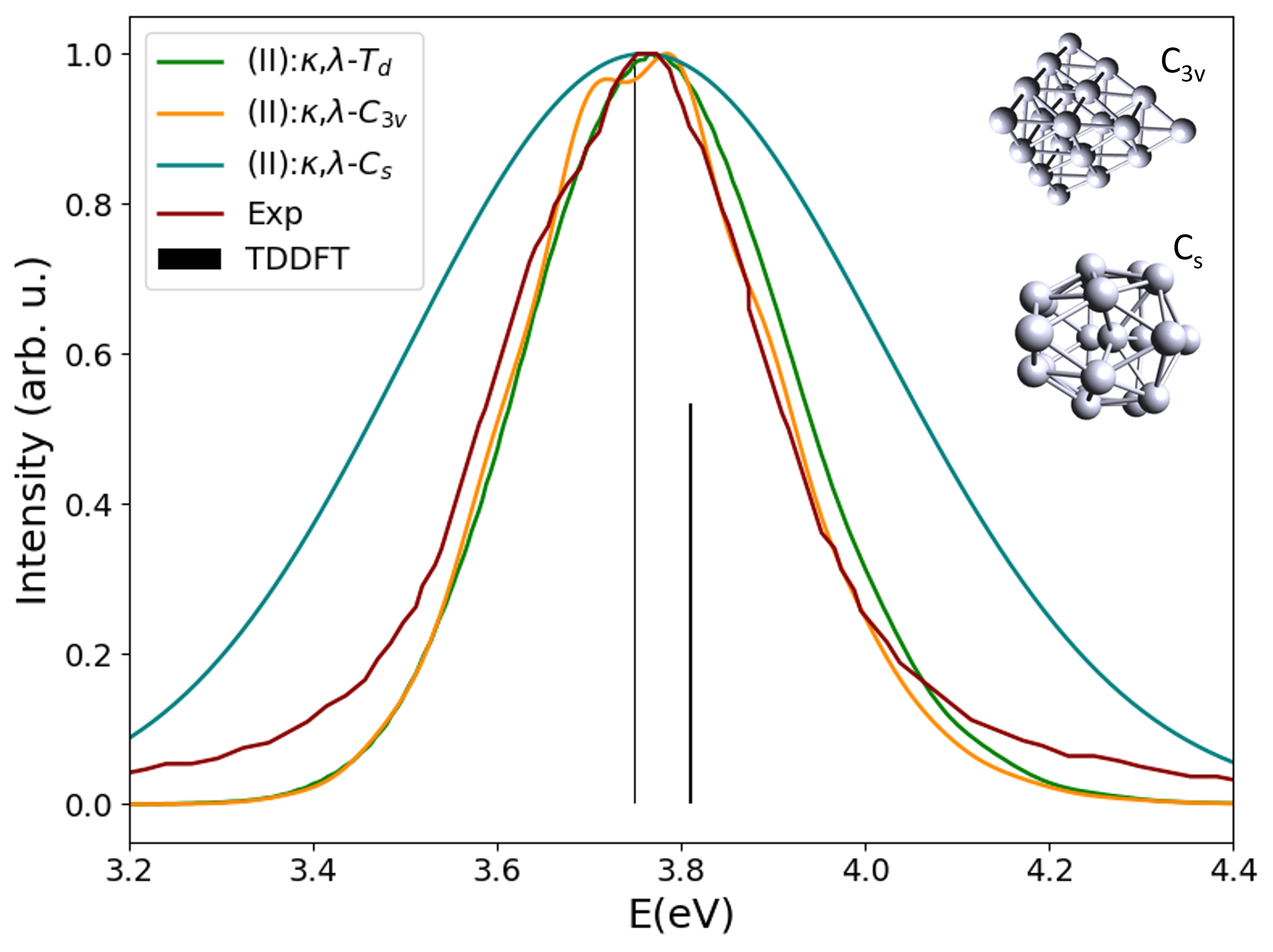}
        \caption{}
        \label{fig:fig7b}
    \end{subfigure}
    \vspace{1.0em} 
    \begin{subfigure}[t]{0.45\textwidth}
        \centering
        \includegraphics[width=\textwidth]{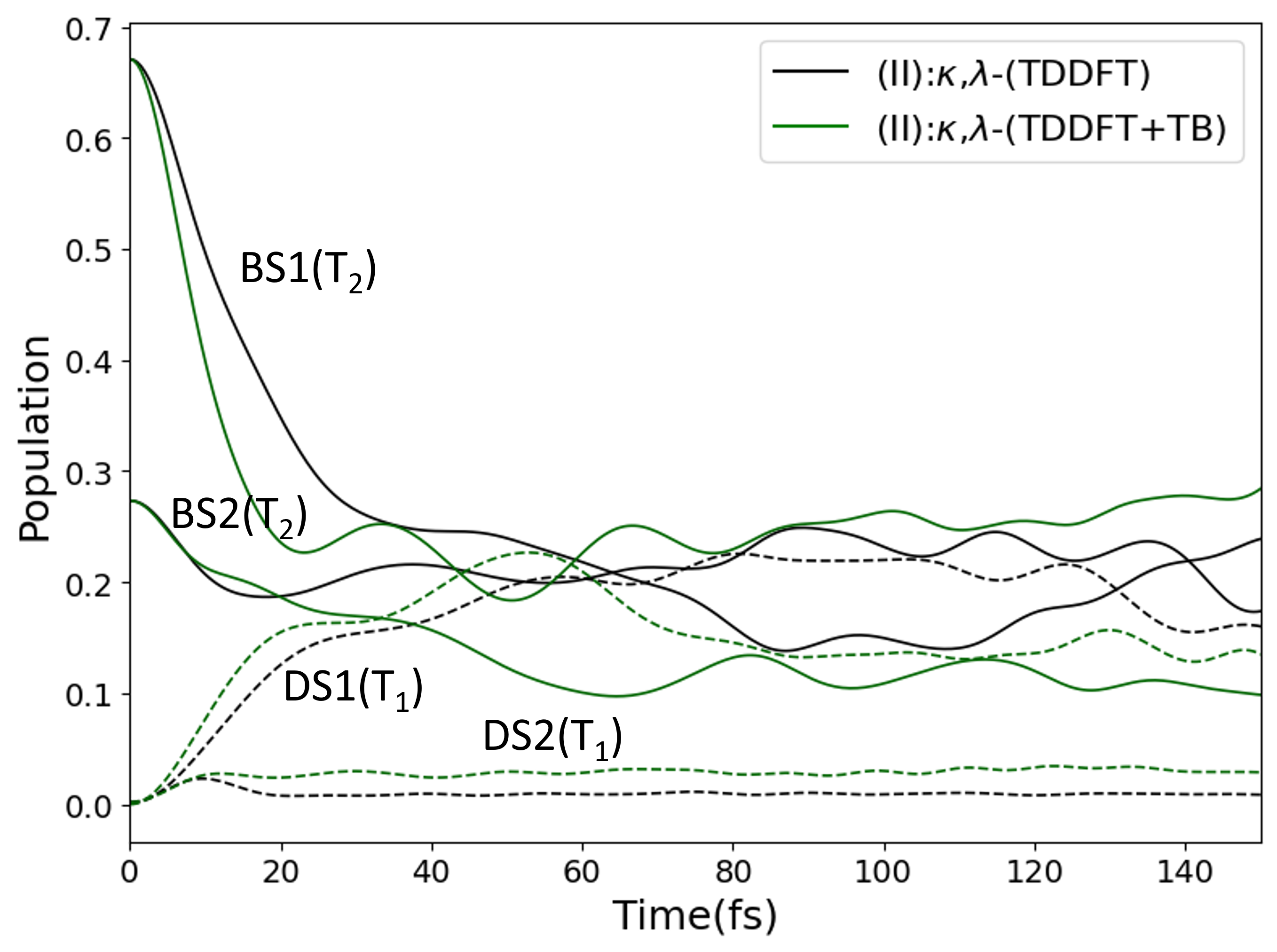}
        \caption{}
        \label{fig:fig7c}    
    \end{subfigure}
        \vspace{1.0em} 
    \begin{subfigure}[t]{0.45\textwidth}
        \centering
        \includegraphics[width=\textwidth]{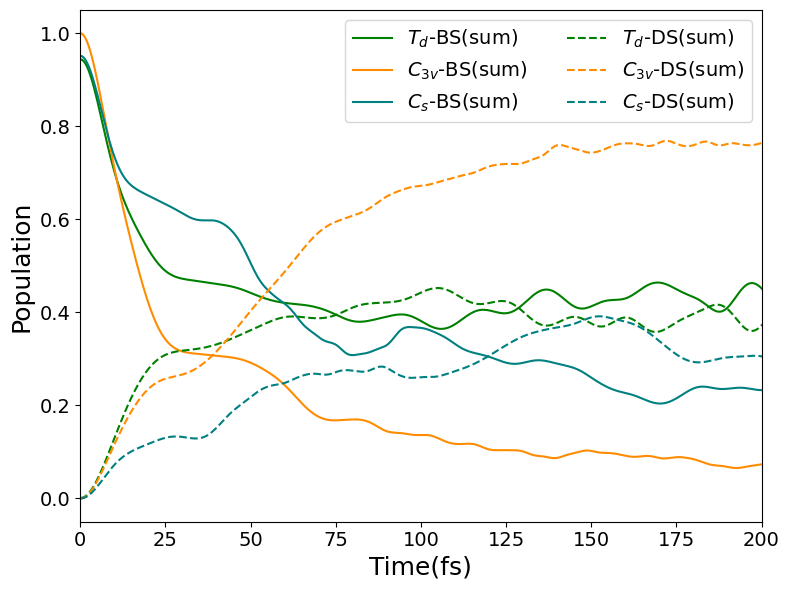}
        \caption{}
        \label{fig:fig7d}    
    \end{subfigure}
   \caption{Absorption spectra (top) and population dynamics (bottom) for selected dark (dashed lines) and bright (solid lines) states of Ag$_{20}$. Panels (a) and (c) correspond to Ag$_{20}$ with $T_d$ symmetry, calculated using the TDDFT and TDDFT+TB methods. Panels (b) and (d) show results for Ag$_{20}$ with $T_{d}$, $C_{3v}$ and $C_s$ symmetries, calculated using the TDDFT+TB method.  'II' refers to the generated $\lambda$ through pipeline~II. The broadening of spectrum is without considering the damping function. The data for the experimental spectrum (red line) are from Ref~\citenum{yu2018optical}.}
    \label{fig:figure7}
\end{figure}

In such systems, the diagonal and off-diagonal terms of the Hamiltonian can effectively transform into one another when the symmetry is reduced from $T_{d}$ to $C_{3V}$ or when different normal modes are involved. To illustrate this, we examined the $Ag_{10}$ cluster by lowering its symmetry from $T_{d}$ to $C_{3V}$ and compared the diagonal coupling terms of the $C_{3V}$ structure with the off-diagonal coupling terms of the $T_{d}$ structure. The subgroup reduction from $T_{d}$ to $C_{3V}$ decomposes the $t_{2}$ irreducible representation into $a_{1}$ and $e$ components (see Tables S2 and S3). Remarkably, for the $C_{3V}$ cluster, the mean absolute value of the intrastate coupling terms, $\overline{\left| \kappa_{a_{1},e}^{(n)} \right|}$ (see Table S16), is nearly identical to the mean absolute value of the interstate coupling terms in the $T_{d}$ point group, $\overline{\left| \lambda_{t_{2},e}^{(n,m)} \right|}$ (see Table S12).
\begin{figure}[ht!]
    \centering
    \begin{subfigure}[t]{0.45\textwidth}
        \centering
        \includegraphics[width=\textwidth]{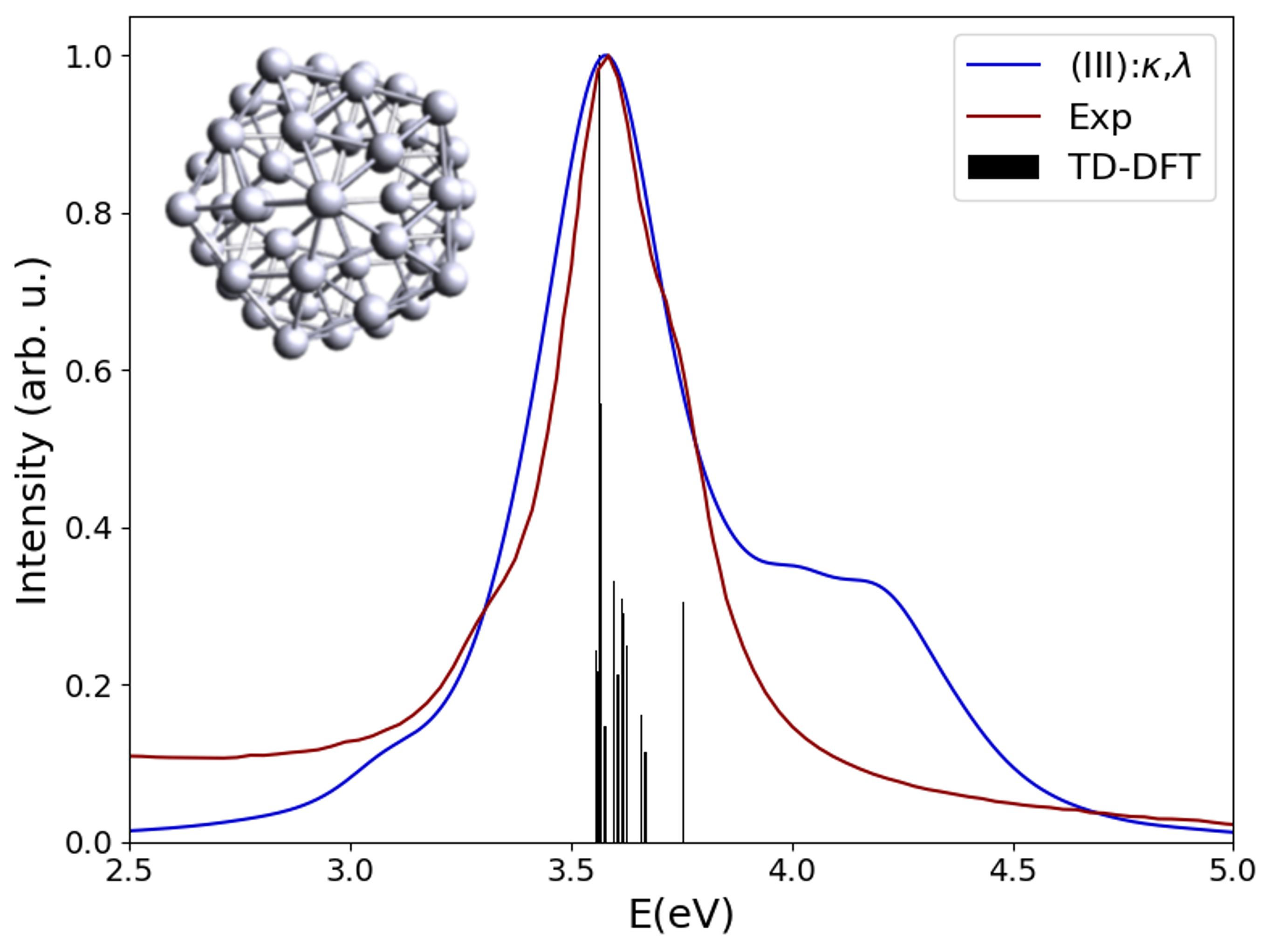}
        \caption{}
        \label{fig:fig8a}
    \end{subfigure}
    \vspace{1.0em} 
        \begin{subfigure}[t]{0.45\textwidth}
        \centering
        \includegraphics[width=\textwidth]{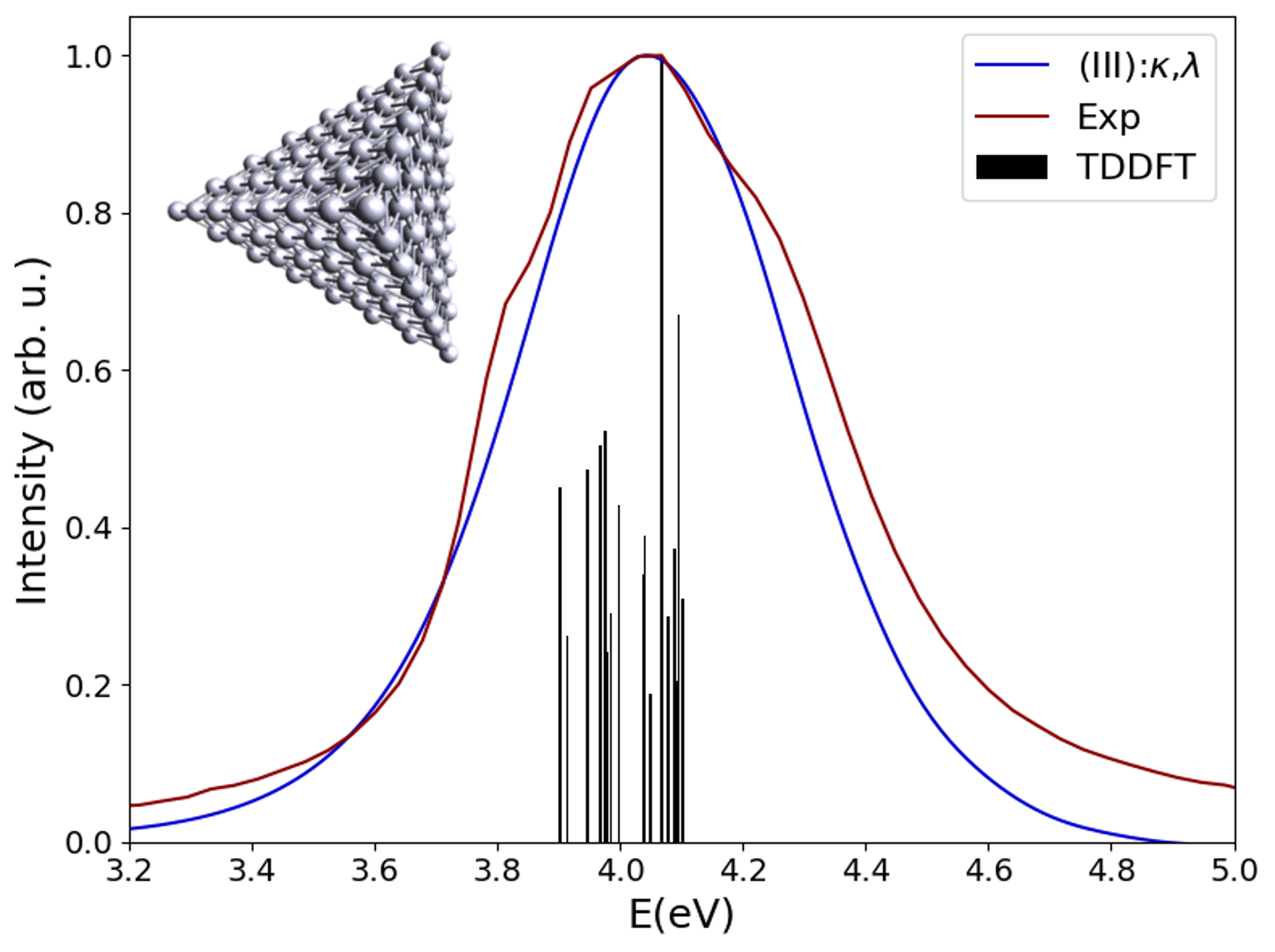}
        \caption{}
        \label{fig:fig8b}
    \end{subfigure}
        \begin{subfigure}[t]{0.45\textwidth}
        \centering
        \includegraphics[width=\textwidth]{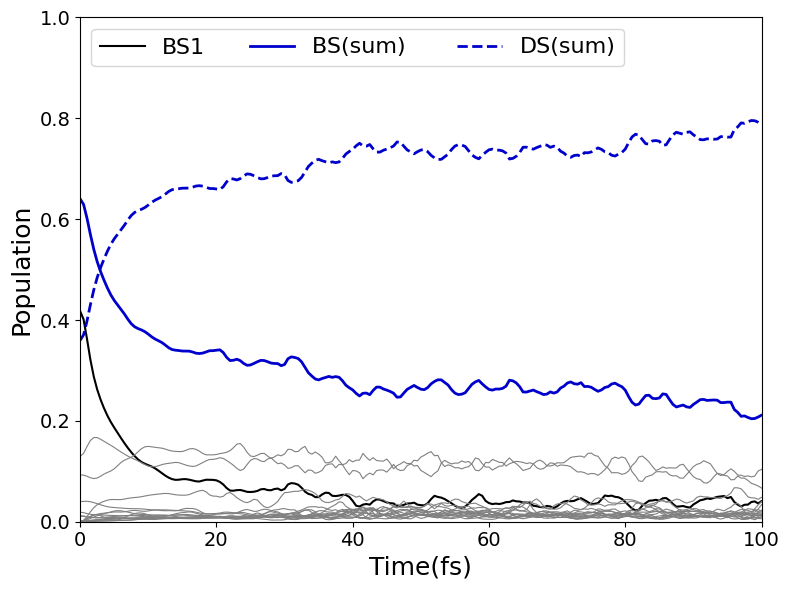}
        \caption{}
        \label{fig:fig8c}
    \end{subfigure}
    \vspace{1.0em} 
        \begin{subfigure}[t]{0.45\textwidth}
        \centering
        \includegraphics[width=\textwidth]{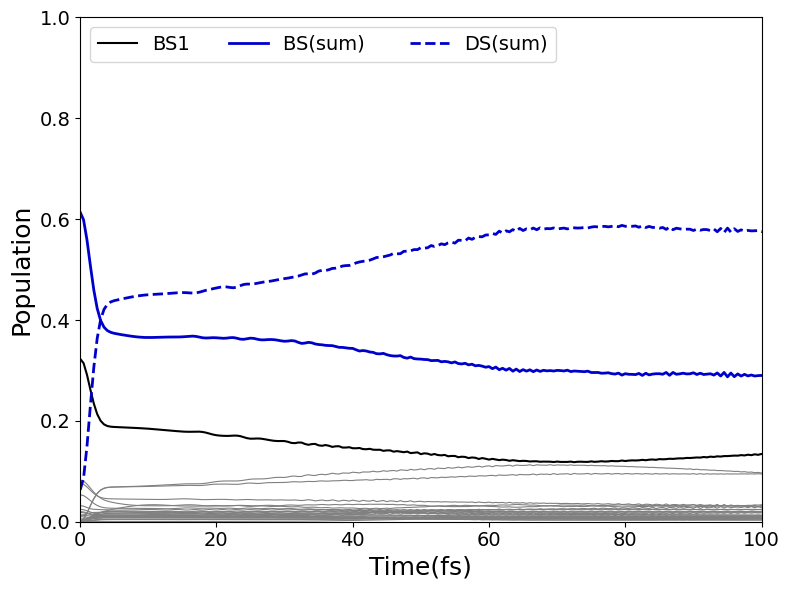}
        \caption{}
        \label{fig:fig8d}
    \end{subfigure}
    \caption{Absorption spectra (top) and population dynamics (bottom) for Ag$_{55}$ $(I_{h})$ (a),(c) and Ag$_{120}$ $(T_{d})$ (b),(d). 'III' refers to the generated $\kappa$ and $\lambda$ through pipeline ~III. The broadening of spectrum is without considering the a damping function. The data for the experimental spectrum (red line) are from Ref~\citenum{schira2019localized}.}
    \label{fig:figure8}
\end{figure}
Therefore, we constructed a dataset based on the $\kappa_i^{(n)}$ values of the $C_{3V}$ structure and used it to estimate the $\lambda_{i}^{(n,m)}$ terms for the corresponding $T_{d}$ structure. As outlined in the previous section, this capability was incorporated into the PyPC framework, enabling the generation of interstate coupling terms from datasets of small- and medium-sized clusters. Furthermore, PyPC includes dedicated modules for analyzing and visualizing the distributions of $\kappa_i^{(n)}$ and $\lambda_{i}^{(n,m)}$, thereby providing deeper insight into their behavior across different symmetry configurations (see Figure \ref{fig:fig2}).

Figures \ref{fig:fig7a} -- \ref{fig:fig7d} present the absorption spectra and population dynamics of Ag$_{20}$. The $\lambda_{i}^{(n,m)}$ terms were determined from the statistical distribution of small- and medium-sized Ag$_{n}$ clusters. Data selection was guided by the symmetry of the vibrational modes and the nature of the transitions (bright or dark).
For the LVC Hamiltonian constructed using TD-DFT+TB $\kappa_i^{(n)}$, the spectral width is broader than that obtained with full TD-DFT. 
However, Ag$_{20}$ ($T_{d}$) spectrum is narrower in comparison to the experimental spectrum (red line in Figure \ref{fig:fig7a}). 
To reproduce the broadening of the experimental spectrum of $Ag_{20}$, the vibronic intrastate parameters for the $T_{d}$, $C_{3v}$, and $C_{s}$ clusters were extracted from pipline~II (see Figure \ref{fig:fig7b}).

For the large cluster, Ag$_{55}$ and Ag$_{120}$, the LVC Hamiltonian was constructed using pipeline~III, with $\kappa_i^{(n)}$ and $\lambda_{i}^{(n,m)}$ parameters generated via the GRCDE method from datasets extracted from smaller clusters. The resulting absorption spectra are compared with experimental measurements \cite{yu2018optical} in Figures \ref{fig:fig8a} and \ref{fig:fig8b}. As illustrated, the model Hamiltonian—comprising over 5,000 parameters—successfully reproduces the experimental spectra with excellent agreement. With increasing cluster size, the absorption spectra exhibit more pronounced broadening, featuring a dominant intense peak that closely mirrors the spectral characteristics of bulk silver.

In Figures \ref{fig:fig8c} and \ref{fig:fig8d}, the population dynamics—summed over all dark states (blue dashed line) and bright states (blue solid line)—display a symmetric pattern of population transfer between these states. Additionally, the dynamics of the bright state with the largest transition dipole moment, BS1 (black solid line), are shown for Ag$_{55}$ and Ag$_{120}$, respectively. As reported in our previous study \cite{asadi2020td}, this state exhibits plasmonic character and undergoes rapid dissipative decay, with lifetimes of less than $\sim20$ fs for Ag$_{55}$ and $\sim10$ fs for Ag$_{120}$.

\section{Conclusions}
The LVC Hamiltonian is a well-established model for studying photo-chemical complex systems and non-adiabatic transitions with multiple electronic and nuclear vibrations, yet it holds emerging, untapped potential in the study of plasmonic cavities made of metallic nano-structures. In this work, we have thus introduced a new python-based platform --- PyPC  tools --- design for the parameterization of a general vibronic coupling model within the framework of Hermitian Hamiltonians. 
Our implementation delivers a black-box methodology capable of: 
\begin{enumerate}
    \item Extracting  LVC parameters --- including vertical excitation energies, gradients, and non-adiabatic couplings --- from first-principle calculations.
    \item Benchmark and extrapolate data obtained from small systems with a generative machine-learning model, thereby generating intra- and interstate coupling parameters for systems that are far larger than those accessible to state-of-the-art first-principles electronic structure methods.
    \item The analysis and determination of the relevant degrees of freedom and the number of modes to include in a particular model to accurately reproduce experimental broadening.
\end{enumerate}
This new implementation enables accurate quantum dynamics simulations of plasmonic nano-particles with $20$ to $120$ atoms. When reaching systems of this size, the explicit modeling based solely on FP data becomes very time-consuming and in some cases impossible. Nonetheless, in this work, we have modeled the non-radiative decay of plasmonic excitations within a Hermitian Hamiltonian framework, employing quantum wavepacket dynamics via the ML-MCTDH method; to demonstrate the capabilities of the approach, we computed the absorption spectra and electronic population dynamics. %
These results illustrate how 
the use of data-driven models --- to replace the electronic-structure calculations --- is a practical alternative and that 
the combination of these new tools enables the modeling of PNCs,  while providing helpful insights into the understanding of plasmonic states. 
Exploring the coupling between emitters and plasmonic nanocavities, both in long- and short-range regimes, is an extension that we are currently pursuing. 

\section{Supporting Information}
The PyPC code is available at \url{https://github.com/Jamshidi-Lab/PyPC}; a detailed explanation of the Generative Regression by Conditional Density Estimation (GRCDE) method used to generate linear vibronic coupling parameters for large silver nanoclusters where first-principles calculations are computationally prohibitive; the ML-MCTDH and tensor tree formats of the wavefunctions for Ag$_{10}$ and Ag$_{55}$; histogram of vibrational frequency distribution of Ag$_{20}$ ; ground state vibrational frequencies of the Ag$_{10}$, Ag$_{20}$, Ag$_{55}$, and Ag$_{56}$; excited state energies and transition dipole moments of the Ag$_{10}$, Ag$_{20}$, Ag$_{55}$, Ag$_{56}$, and Ag$_{120}$; off-diagonal linear coupling constants of Ag$_{10}$; on-diagonal linear coupling constants of Ag$_{20}$ in TD-DFT and TD-DFT+TB level of theory.

\begin{acknowledgement}
Z.J. would like to thank the Alexander von Humboldt’s research foundation and
    scientific research grant G4040605 for financial support.

\end{acknowledgement}

\bibliography{Reference}
\end{document}